\documentclass[a4paper,12pt]{article}
\usepackage{graphicx}
\usepackage[lflt]{floatflt}
\usepackage{amsbsy}
\usepackage{amsbsy}
\usepackage{amsmath}
\usepackage{amssymb}
\usepackage{amsfonts}

\begin{document}

\begin{center}
{ \Large
\textbf{Gravitational Lensing in presence of Plasma:\\Strong Lens Systems, Black Hole Lensing and Shadow}
}\\[20pt]

{ \large
Gennady S. Bisnovatyi-Kogan$^{1,2}$ and Oleg Yu. Tsupko$^{1}$
}\\[20pt]

$^{1}$Space Research Institute of Russian Academy of Sciences,\\ Profsoyuznaya 84/32, Moscow 117997, Russia;\\
$^{2}$National Research Nuclear University MEPhI (Moscow Engineering\\Physics Institute), Kashirskoe Shosse 31, Moscow 115409, Russia\\[20pt]

emails: gkogan@iki.rssi.ru, tsupko@iki.rssi.ru\\[20pt]

Published in 2017\\[10pt]

\end{center}

\abstract{ In this article, we present an overview of the new developments in problems of the plasma influence on the effects of gravitational lensing, complemented by pieces of new material and relevant discussions. Deflection of light in the presence of gravity and plasma is determined by a complex combination of various physical phenomena: gravity, dispersion, refraction. In particular, the gravitational deflection itself, in a homogeneous plasma without refraction, differs from the vacuum one and depends on the frequency of the photon. In an inhomogeneous plasma, chromatic refraction also takes place. We describe chromatic effects in strong lens systems including a shift of angular position of image and a change in magnification. We also investigate high-order images that arise when lensing on a black hole surrounded by homogeneous plasma. The recent results of analytical studies of the effect of plasma on the shadow of the Schwarzschild and Kerr black holes are presented.}

\section{Introduction}

\subsection{Refraction}


Gravitational lensing theory describes a wide range of phenomena connected with deflection of light by gravity. Gravitational lensing includes many effects:
a change of apparent angular position of the source, multiple imaging, magnification (change of the flux), distortion (change of the shape), time delay. Now
gravitational lensing is a powerful astrophysical tool for
investigations of distant objects, a distribution of dark
matter and large scale structure, the cosmic microwave
background, discovery of planet and checking of the General
Relativity, see some reviews, classical and interesting recent works \cite{GL, GL2, Perlick2004review, Mao2012-review, BartelmannSchneider-review,
Bartelmann-review, Wambsganss-review, Hoekstra-review, Refsdal1964a, Refsdal1964b, Clark1972, Byalko1969, Mao1991, Nucita-Paolis-2014, Zhdanov2010, Zhdanov2011,
Er-Li-Mao-Cao-2013, Clowe2006, Kayser1986, WambsganssPaczynski1991, Zakharov-UFN, Cherep01, Cherep02}. In present paper we discuss how presence of plasma may influence various effects of gravitational lensing.

In the outer space, the rays of light travel through the plasma. In plasma, photons undergo various effects, such as absorption, scattering and refraction.
For gravitational lensing, which is usually used in the approximation of geometric optics, the main interest is a change of the angle of deflection of a light
ray. Due to dispersion properties of plasma, we may expect also that chromatic effects will arise.

\begin{figure}[h]
	\centering
	\includegraphics[width=14cm]{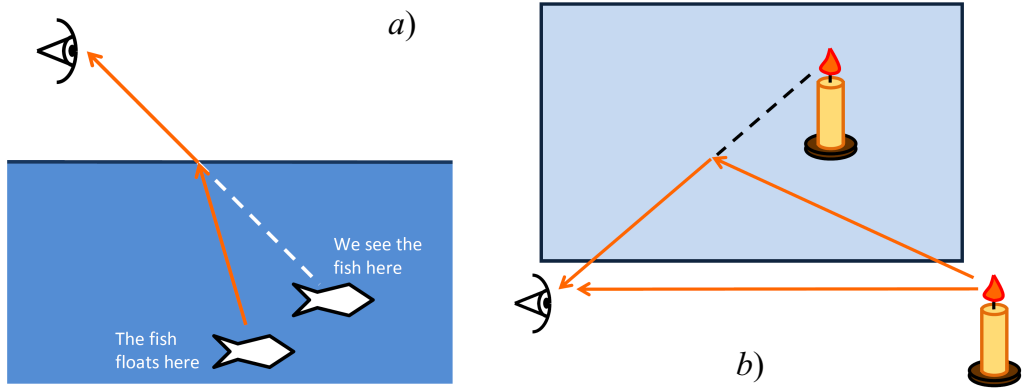}
	\caption{Effects similar to the effects of gravitational lensing but arising from the non-homogeneity of the medium: (\textbf{a}) Change in the visible position of the object due to the refraction of light at the interface between two media. (\textbf{b}) The emergence of two images of the same object when placed near a mirror.}
	\label{figure-fish-candle}
\end{figure}

It is widely known that light rays in a transparent, inhomogeneous medium propagate along curve trajectories \cite{LL8, Zhelezn}. This phenomenon is called refraction and is well
 known from everyday life. For example, due to the refractive effect, the image when viewed through the optical lens differs from the true one, a spoon in the
 glass of water seems to be broken, and a depth of the pond seems to be less than in reality. The bending of the light rays due to refraction is not related to
 relativity or gravity and occurs only if the medium is optically non-homogeneous (Fig. \ref{figure-fish-candle}).

The simplest way to add plasma to the problem of the gravitational lensing is to consider cases, when  both deflection angles,
due to gravity and due to refraction in the plasma, are small, when these angles can be calculated separately from each other.
To calculate the gravitational angle of deflection, it is sufficient to use the linearized theory of gravitation, i.e. to use
the approximate Einstein formula for the deflection angle. To calculate the refraction, we can assume that the refractive index
of the plasma differs only slightly from unity (vacuum).

In this approximation, the joint action of gravitation and refraction was investigated from the 1960's, with reference to the
propagation of radio signals in the solar corona. The rays of light, passing close to the Sun, are deflected both by solar gravity
and by plasma in the solar corona. The corresponding angles of deflection were calculated by Muhleman et al \cite{Muhleman1970},
\cite{Muhleman1966}, see also \cite{zadachnik}. The deflection due to refraction in this case is chromatic, i.e. it depends on the
frequency of the photon. The same approximation was used by Bliokh and Minakov \cite{BliokhMinakov} who first considered this problem
in the context of gravitational lensing. They investigated a point gravitational lens surrounded by an inhomogeneous plasma with a
power law density distribution.\\

\subsection{General theory of the geometrical optics in a medium, in presence of gravity}

For a more detailed study of the effects of gravitational lensing in a plasma, a more rigorous approach is required. First of all,
black holes and other compact objects can also be surrounded by a plasma. To describe the rays passing near such a compact object,
the linearized theory of gravity is insufficient. The second interesting question is whether the gravitational deflection itself
is changed in the presence of a medium around a gravitating body, in comparison with a vacuum case. We should also not forget that
in addition to a point lens, there are many complex lens models. It turns out that more rigorous treatment of the subject reveals
more complicated behaviour and new chromatic effects.

In order to deal with the issues mentioned above, it is necessary to use a theory that allows simultaneously to describe both the
curvature of the spacetime (gravity) and the presence of the medium. The first self-consistent approach to the light propagation in
 the gravitational field in presence of a medium was
developed in the classical book by Synge \cite{Synge}, for subsequent discussions see \cite{Bicak1975, Krikorian1985, Krikorian1999},
see also work of Kulsrud and Loeb \cite{Kulsrud-Loeb}. The comprehensive review of the general relativistic ray optics
in media is presented in the monograph
by Perlick \cite{Perlick2000}. In particular, general formulas for the light deflection angle in the Schwarzschild and Kerr metrics
(in the equatorial plane) in the presence of a
spherically distributed plasma have been obtained in the integral form \cite{Perlick2000}, see also \cite{TsBK2013}, \cite{BKTs2015}.
For propagation of electromagnetic waves in magnetized plasma see series of papers of Breuer and Ehlers
\cite{BreuerEhlers1980, BreuerEhlers1981, BreuerEhlers1981-AA}, and of Broderick and Blandford
\cite{Brod-Blandford-MNRAS-1, Brod-Blandford-MNRAS-2, Brod-Blandford-ASS}.

Detailed investigation of physical properties and different situations of gravitational lensing in plasma have been started in
series of papers of Bisnovatyi-Kogan and Tsupko, see Refs.~\cite{BKTs2009}, \cite{BKTs2010}, \cite{TsBK2013}, for review see Ref.~\cite{BKTs2015}.
We wrote down the Synge's equations for the Schwarzschild metric and investigated various situations and models of lensing. In particular, it
has been shown that gravitational deflection itself is different from vacuum case, and is chromatic in plasma, even in the case when plasma is
homogeneous and refractive deflection is absent \cite{BKTs2009}, \cite{BKTs2010}. In non-homogeneous plasma both,  chromatic
gravitational deflection and chromatic refractive deflection are present. So, the presence of plasma always makes gravitational lensing chromatic.
In particular, it leads to a difference
in the angular positions of the same image at different
wavelengths. Gravitational lensing in plasma was also investigated at large deflection angles, when the light rays pass very close to the black
hole and so called high-order (relativistic) images are formed \cite{TsBK2013}. Subsequently, investigations of lensing in plasma have been
continued by many authors \cite{Morozova2013, Mao2014, Perlick2015, Rogers2015, Rogers2016-arxiv,
 Rogers2017-MNRAS, Rogers2017-Universe, Hakimov2016, Liu2016-Kerr, PerlickTsupko2017},
 mostly with using the same formalism, see details below.\\

\subsection{Strong lens systems}

\begin{figure}[h]
	\centering
	\includegraphics[width=15cm]{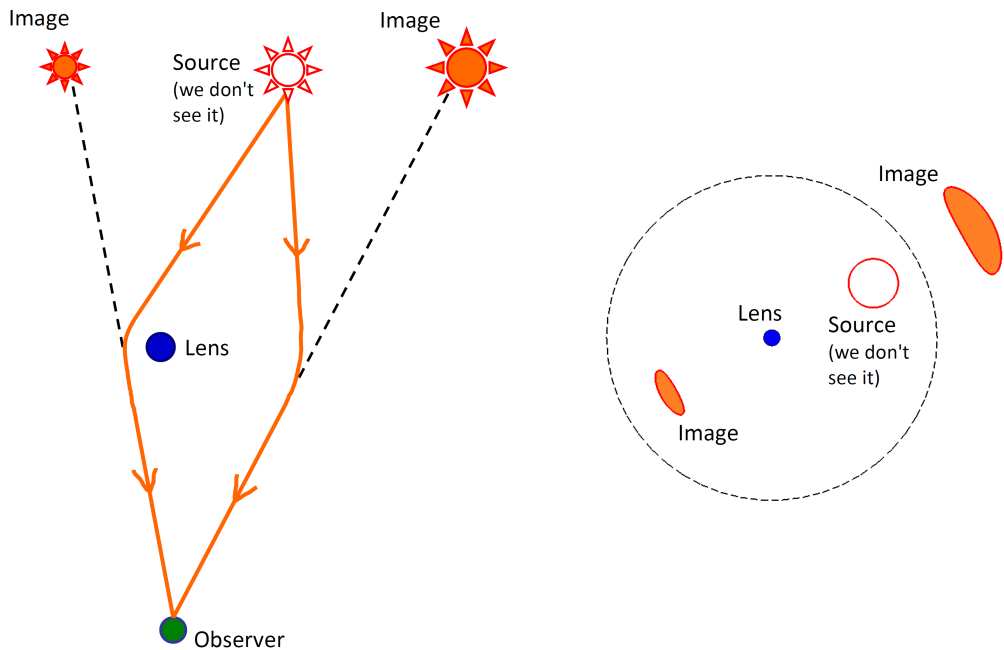}
	\caption{Formation of multiple images of the same source by gravitational lensing in vacuum. The left picture shows a general
geometry of the problem and trajectories of the light rays. The right picture demonstrates the positions of images at the observer's
sky. A dashed circle is the Einstein-Chwolson ring.}
	\label{figure-multiple}
\end{figure}

One of the most beautiful effects of gravitational lensing is appearance of multiple images of the same source. This effect is easily demonstrated
in the simplest case of lensing by point mass (Fig. \ref{figure-multiple}). In the case of a strong lens system the distribution of matter in lens
is more complicated than just point mass,
with essential deviations from spherical, or any types of symmetry.
 Nevertheless, the weak deflection approximation still can be used, meaning
that each small portion of matter deflects light in accordance with the approximate Einstein formula.

At the present level of development of observations, chromatic effects of lensing in a plasma might be observed
 only in a strong lens system\footnote{We call here as a ``strong lens system'' a lens with several visible images of the same source, but the deflection angle remains to be small. This situation should not be confused with ``lensing in strong deflection (or strong gravity) regime'' when the deflection angles of photons are not small and approximate Einstein formula can not be applied.}. This means that the lens is a galaxy or a cluster of galaxies,
 and there are several images of the same distant source, for example, quasar. Due to the presence of plasma, when observing images
 in different wavelengths, we expect to see different angular positions. By measuring the difference in  angular positions, we can
 obtain information about the distribution of the plasma in the lens. Calculation of the deflection angles for different mass and plasma
 distributions in the lens is done in the work of Bisnovatyi-Kogan and Tsupko \cite{BKTs2010}. For the case of a homogeneous plasma, positions
 and magnifications of images were  calculated analytically. Modelling of plasma effects in strong lens systems, similar to Einstein
 Cross, was performed in the paper of Er and Mao \cite{Mao2014}. Authors have found that effects of plasma are possible to be detected
 in the low frequency radio observations.\\

\subsection{Black hole lensing}

Unlike stars and galaxies, in the case of lensing by a black hole, the impact parameter of a photon can be of the order of the Schwarzschild
radius. In this case, the action of gravity is strong, and the deflection angle of the photon is not small. Linearized theory of gravity is not
sufficient, and the deflection angle can not be calculated by Einstein's formula. In a vacuum, it is necessary to use the exact equations of
photon motion, which lead to an integral expression for the deflection angle. In the case of plasma, it is necessary to obtain exact equations
of the photon motion, using either the Singe's theory \cite{Synge}, or the Perlick's method \cite{Perlick2000}, which again leads to an
expression for the deflection angle in the integral form.

\begin{figure}[h]
	\centering
	\includegraphics[width=14cm]{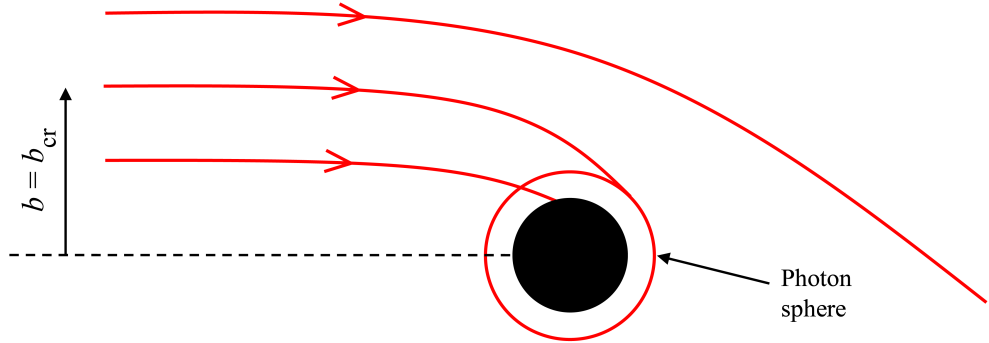}
	\caption{Schematic representation of trajectories of light rays incident from infinity toward a black hole with different impact parameters. Photons with $b<b_{cr}$ are absorbed by the black hole. The photons within $b>b_{cr}$ after approaching black hole flies off to infinity. Light rays with $b=b_{cr}$ are travelling to a photon sphere which contains unstable photon circular orbits. For the Schwarzschild space-time the Schwarzschild radius equals to $2M$, the photon sphere radius equals to $3M$, and critical impact parameter equals to $b_{cr} = 3\sqrt{3}M$. In case $b \gg 3\sqrt{3} M$, the photon deflection angle is small and can be calculated by approximate Einstein formula $\hat{\alpha} = 4M/b$.}
	\label{figure-diff-impact}
\end{figure}

\begin{figure}[h]
	\centering
	\includegraphics[width=14cm]{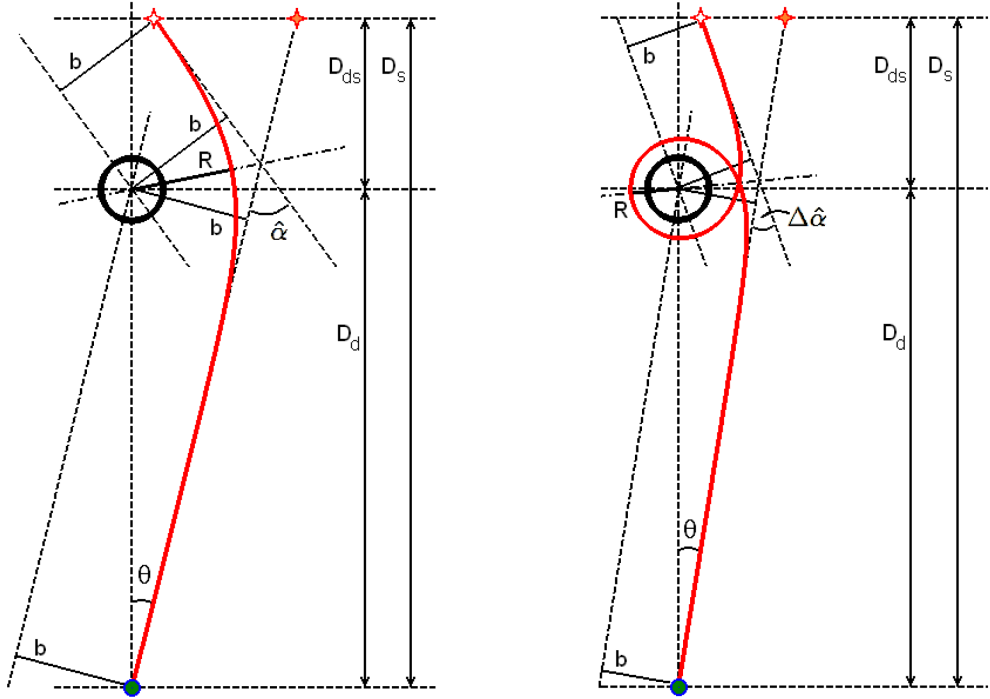}
	\caption{Scheme of formation of primary (left) and high-order (right) images of source in case of gravitational lensing by black hole in vacuum. Trajectories of
		light rays are calculated in the Schwarzschild metric. Light ray from the source deflected by the
		point-mass gravitational lens goes to the observer. The observer sees the
		image of the source at angular position $\theta$, which is different from the real source position. $R$ is the closest point
		of trajectory to gravitating center, it is usually referred as the distance of the closest approach, $b$ is the
		impact parameter of the photon, $D_d$ is the distance between observer and lens, $D_s$ is the distance between
		observer and source, $D_{ds}$ is the distance between lens and source. In case of high-order image (right), the light ray from the source is deflected by angle $\hat{\alpha} = 2 \pi +
\Delta \hat{\alpha}$, and the impact parameter $b$ is close to its critical value $b_{cr}$. }
	\label{figure-prim-relat}
\end{figure}

\begin{figure}[h]
	\centering
	\includegraphics[width=9cm]{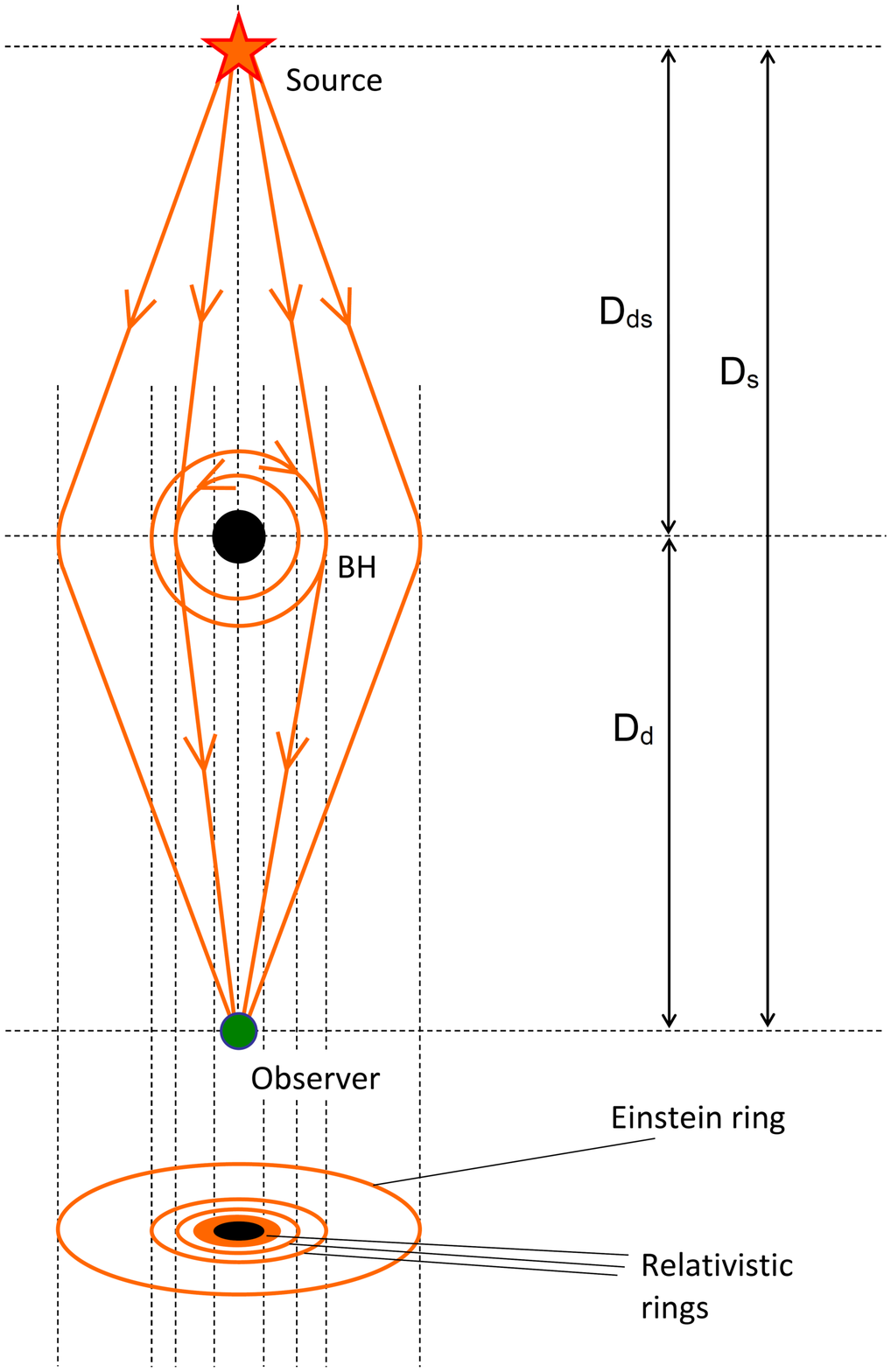}
	\caption{Scheme of formation of Einstein ring and relativistic rings by black hole lensing.}
	\label{figure-rings}
\end{figure}

There is a critical value of the impact parameter, which divides the incident photons into two types (Fig. \ref{figure-diff-impact}). Photons, in which
the impact parameter is less than critical, are absorbed by a black hole. The photons with the impact parameter greater than the critical, after
approaching the black hole, again fly off to infinity. In this case, if the impact parameter is close to the critical one, the photons can make
one or several revolutions around a black hole before flying off to infinity. Such photons give rise to an infinite sequence of images on both
sides of the black hole, these are so-called high-order images or relativistic images. Thus, when lensing on a black hole, there are, first,
two 'ordinary' images of the source (formed by the photons with $\hat{\alpha} < 2\pi$), and secondly, two infinite sequences of high-order images (formed by the photons with $\hat{\alpha} > 2\pi$), see Fig. \ref{figure-prim-relat}.
Note that in case of the perfect alignment of the source, the black hole and the observer, two ordinary images merge in one image -- the Einstein ring, and there is the infinite sequence of relativistic rings, see Fig. \ref{figure-rings}.

The calculation of relativistic images can be made using the exact deflection angle in the lens equation. Exact deflection angle of photon
 moving in vacuum in the Schwarzschild metric was calculated in paper of Darwin \cite{Darwin1959}, it has a form of the integral, and can
 be expressed via elliptic integrals \cite{Darwin1959, BKTs2008}. By using the exact expression for the deflection angle, the properties of relativistic images had been calculated in paper of Virbhadra and Ellis \cite{Virbhadra2000}, see also paper \cite{Virbhadra2009}. The
  exact gravitational lens equation in spherically symmetric and static space-times was considered by Perlick \cite{Perlick2004a}.

The investigation of relativistic images is simplified if we use the so-called strong deflection approximation which is opposite to weak
deflection approximation. This approximation means that the deflection angle is much greater than unity. When the impact parameter approaches
the critical value, the deflection angle logarithmically diverges, formally tending to infinity. The logarithmically divergent expression for
the photon deflection angle in vacuum in the strong deflection limit was obtained by Darwin for the Schwarzschild metric \cite{Darwin1959},
see also \cite{Bozza2001}.

Strong deflection limit provides very high accuracy for photons that have made one or more turns, so it is convenient to use it to calculate the properties of relativistic images. Analytical calculation of the positions and magnifications of relativistic images was made in the article of the Bozza et al \cite{Bozza2001}. Later Bozza \cite{Bozza2002} has derived analytical expression for the deflection angle in general form, for generic spherically symmetric space-time. In particular, it was shown that for any spherically symmetric space-time the deflection angle diverges logarithmically when the impact parameter approaches its critical value. Relativistic rings (Fig.\ref{figure-rings}) were investigated in details in work \cite{BKTs2008}.

The subject of gravitational lensing beyond weak deflection approximation becomes very popular, because in the strong field regime it is possible to
differ and compare different theories of gravity, see, for example, Refs.~\cite{BKTs2008, Perlick2004review, PerlickMG, Perlick2002, Virbhadra2001,
Virbhadra2008, Bozza2008, Bozza2010, Eiroa2002, IyerPetters, EiroaSendra, Cite1, Cite2, Cite3, Cite4, Cite5, Cite6, Cite7}. Nevertheless,
there are no big chances so far to observe the high-order images.

Similarly to photons, a massive particle incident from infinity to the gravitating center, for certain values of the parameters, can also make several revolutions around the central body and fly away to infinity. Radius of unstable circular orbits was investigated in works of Zakharov \cite{Zakharov-1988}, \cite{Zakharov-1994}. Recently, analytical expressions for deflection angles of massive particles in the Schwarzschild metric in strong deflection limit were derived for the first
time by Tsupko \cite{Tsupko2014}. The same problem have been considered later in paper of Liu, Yang and Jia \cite{Liu2016-massive}, in the context of
 gravitational lensing of massive particles.

Investigation of gravitational lensing in plasma beyond weak deflection approximation is made in Ref.~\cite{TsBK2013}. The angle
of deflection of a photon in the Schwarzschild field and in a homogeneous plasma was derived. The positions and intensities of relativistic
images were analytically calculated. It is shown that the presence of the uniform plasma increases the angular size of relativistic rings
or the angular separation of point images from the gravitating center. The presence of the uniform plasma increases also a magnification
of relativistic images \cite{TsBK2013}. This consideration have been recently extended to Kerr spacetime by Liu, Ding and Jing \cite{Liu2016-Kerr}.

Complicated behaviour of light rays near compact objects was demonstrated in series of papers of Rogers
\cite{Rogers2015, Rogers2016-arxiv, Rogers2017-MNRAS, Rogers2017-Universe}. Trajectories of light rays are investigated in spherically
symmetric metric with different power-law plasma distribution. As a relevant physical example the propagation of radiation emitted from
a compact object sheathed in a dense plasma had been examined \cite{Rogers2015}. The plasma distribution significantly affects the appearance
of the compact object for a distant observer, by changing the portion of the surface that is visible as a function of frequency. Rogers also
 investigated stable circular orbits of light rays in non-homogeneous plasma distribution \cite{Rogers2017-Universe}. Bound orbits of light rays in the homogeneous plasma case were discussed in \cite{Kulsrud-Loeb} and \cite{TsBK2013}.\\

\subsection{Lensing by Kerr black hole in presence of plasma}

The deflection angle for a light ray moving near a Kerr black hole surrounded by a plasma was first derived by Perlick in the form of the
integral \cite{Perlick2000}. The derivation is made for the equatorial plane of a black hole, and spherically symmetric distribution of the
plasma. For the non-equatorial motion of a photon, the deflection angle is derived in the article by Morozova, Ahmedov and Tursunov \cite{Morozova2013},
in the weak deflection approximation. These authors generalize the method used in the paper \cite{BKTs2010} for the Schwarzschild metric to the case
of a homogeneous plasma and a weakly rotating black hole. As mentioned, relativistic images in lensing on a Kerr black hole surrounded by a plasma
are investigated in the article \cite{Liu2016-Kerr}. The equations of motion for an arbitrary distance of a photon from a black hole, an arbitrary
ray inclination, and an arbitrary Kerr parameter are obtained in the article of Perlick and Tsupko \cite{PerlickTsupko2017}. The article shows that
 the separation of variables in the Hamilton-Jacobi equation, the existence of the Carter constant generalized for the plasma, and complete integrability of the equations of motion are possible only for specific plasma distributions around the Kerr black hole.\\

\subsection{Black hole shadow}

In close connection with lensing by a black hole is a topic of a black hole shadow. It is expected that a distant observer should 'observe' the black
hole as a dark spot(on the background of bright sources) in the sky which is known as the 'shadow'.

This topic is now becoming very popular due to the appearance of projects to observe the shadow of a supermassive black hole in the center of our Galaxy.
Observing the shadow of black hole is both very challenging and very difficult task due to many effects involved. For the black hole at the center of our Galaxy,
the size of the shadow is about 53 $\mu$as (which approximately corresponds to the size of a grapefruit on the Moon if observed from the Earth).
 At present, two projects are under way to observe this shadow which
would give an important information about the properties of the compact object at the center of
our Galaxy. These projects, which are going to use (sub)millimeter VLBI
observations with radio telescopes distributed over the Earth, are called as the
Event Horizon Telescope (http://eventhorizontelescope.org) and the
BlackHoleCam (http://blackholecam.org).

For a non-rotating black hole, the shadow is a circular disk at the sky. For a Schwarzschild black hole the angular diameter of the shadow
was calculated, as a function of the mass of the black hole, and of the coordinate distance from the black hole to the
observer by Synge \cite{Synge1966}. For the Kerr black hole, the shadow becomes oblate and deformed. The shape of the
shadow of a Kerr black hole, for an observer at infinity,
was calculated by Bardeen \cite{Bardeen1973}. In the papers \cite{GrenzebachPerlickLaemmerzahl2014},
\cite{GrenzebachPerlickLaemmerzahl2015}, the size and the
shape of the shadow were calculated for the whole class
of Pleba{\'n}ski-Demia{\'n}ski spacetimes (which includes the Kerr black hole as a
special case). Calculations are performed for the observer at an arbitrary position outside of the horizon of
the black hole. Numerous analytical investigations and numerical simulations of the shadow are presented in literature
(for example, see \cite{FalckeMeliaAgol2000, JamesTunzelmannFranklinThorne2015, FrolovZelnikov2011, Luminet1979, Chandra,
Konoplya2016a, Konoplya2016b, Johannsen2016,  HiokiMaeda2009, LiBambi2014, TsukamotoLiBambi2014, Takahashi2004, YangLi2016, Abdujabbarov2015,
Bambi2013, Dymnikova1986, Zakharov-1988, Zakharov-1994, Zakharov-Paolis-2005-NewAstronomy, Zakharov-Paolis-2005-AA, Zakharov-Paolis-2012, Zakharov2014, Tsupko2017, Ahmedov-add00, Ahmedov-add01, Ahmedov-add02, blackholecam-review}).

Influence of a matter on the shadow is usually investigated by numerical calculation. In particular,
Falcke, Melia and Agol \cite{FalckeMeliaAgol2000} have numerically simulated
the visual appearance of the black hole at the center of
our Galaxy, assuming that it is a Kerr black hole, with
scattering and the presence of emission regions between
the observer and the black hole taken into account (at
0.6 and 1.3 mm wavelengths). Sophisticated ray tracing
programs have been written for producing realistic
images of a black hole surrounded by an accretion disk,
e.g. for the movie \textit{Interstellar}, see details in \cite{JamesTunzelmannFranklinThorne2015}.

The first attempt of analytical investigation of plasma
influence on the shadow size, in the frame of geometrical optics, taking into
account effects of general relativity and plasma presence, was performed in paper \cite{Perlick2015}. The
Synge's formula was generalized to the case of a
spherically symmetric and static plasma distribution on
a spherically symmetric and static spacetime. As particular
examples, the results for a Schwarzschild black hole and an Ellis
wormhole have worked out.

Recently, the shadow of a Kerr black hole under the influence of a plasma have been investigated analytically
\cite{PerlickTsupko2017}. An analytical formula for the boundary curve of the
shadow on the sky for an arbitrarily situated observer has been derived.\\

\subsection{Additional notes about propagation of light rays in medium} \label{subsection-notes}

Plasma has unique dispersive properties, that leads to interesting analogy between motion
of photon in homogeneous plasma and massive particle in vacuum. In the paper of Kulsrud
and Loeb \cite{Kulsrud-Loeb} (see also papers of Broderick and Blandford \cite{Brod-Blandford-MNRAS-1}, \cite{Brod-Blandford-MNRAS-2}, \cite{Brod-Blandford-ASS}) it was shown that in
the homogeneous plasma the photon wave packet moves like a particle with a velocity equal
to the group velocity of the wave packet, with a mass equal to the plasma frequency (multiplied by $\hbar$), and with an energy equal to the photon energy \footnote{This is true for the massive particle energy and the photon frequency at infinity, and also for static observer at another locations because the particle energy and the photon frequency have the same dependence on the space coordinates arising due to the gravitational field \cite{MTW}.}. Therefore trajectories of light rays moving in homogeneous plasma can be investigated with using results for massive particles moving in vacuum. For example, chromatic deflection of light in homogeneous plasma can be explained by chromatic group velocity of photons (photons of different frequencies have different speeds, and thus their geodesics are bent by gravity in different ways), see also Subsection \ref{subsection-hom-plasma}. Note that this analogy is applicable only for homogeneous plasma, and does not work for medium with arbitrary refraction index, see below.

As for massive particle in vacuum, bound elliptic orbits
of the photons in homogeneous plasma are also possible.
Gravitational binding of the photon in homogeneous
plasma was discussed in the paper of Kulsrud and
Loeb \cite{Kulsrud-Loeb} and in Section VI of paper of Tsupko and Bisnovatyi-Kogan \cite{TsBK2013}. For non-homogeneous plasma distributions, stable orbits of light rays were investigated for the first time by Rogers \cite{Rogers2017-Universe}. Variety of orbits including circular and periodic orbits are revealed, for beautiful illustrations see \cite{Rogers2017-Universe}.

The main properties of the gravitational deflection in different media can be formulated as \cite{BKTs2009}:

(i) In vacuum the gravitational deflection is achromatic.

(ii) If a medium is homogeneous and not dispersive (the refractive index $n$ is just a constant, it does not depend either on the space coordinates or on the photon frequency), the gravitational
deflection angle is the same as in vacuum. Nevertheless, the group speed is smaller than light speed in vacuum.

(iii) If the medium is homogeneous but dispersive (the refractive index does not depend on the space
coordinates but depends on the photon frequency), the gravitational deflection angle is different from the vacuum
case and depends on the photon frequency. For example: plasma.

(iv) If medium is non-homogeneous (the refractive index depends on the space coordinates), we will have
also the refractive deflection. If medium is non-homogeneous and dispersive, the refractive deflection depends
on the photon frequency.\\

The present paper is organized as follows. In next section we consider a calculation of the photon deflection
angle in presence of both gravity and plasma. In Section 3 we discuss gravitational lensing in plasma in case of
strong lens systems with several observed images. In Section 4 we consider black hole lensing, when the deflection
angles can be large and relativistic images can arise. In Section 5 and 6 the influence of plasma on the black
hole shadow is described.\\


\section{Deflection angle in presence of plasma in \\ Schwarzschild metric}

\subsection{Exact expression for the deflection angle}

Let us consider a spherically symmetric and static metric

\begin{equation}\label{eq:g}
ds^2 = g_{ik}dx^{i} dx^{k}
= - A(r) dt^2 + B(r) dr^2 +
D(r) \big( d \vartheta ^2 + \mathrm{sin} ^2 \vartheta \,
d \varphi ^2 \big) \, ,
\end{equation}
where metric coefficients $A(r)$, $B(r)$ and $D(r)$ are positive. Throughout this paper we use $G=c=1$, except Section 6.

We assume that the spacetime is filled with a static non-magnetized
cold\footnote{We assume that typical size of considered systems is much larger than Debye radius, and therefore temperature effects in dispersion relations are neglected.} inhomogeneous plasma whose electron plasma frequency $\omega_e$ is a function
of the radius coordinate only,
\begin{equation} \label{eq:omega-e}
\omega_e^2 = \frac{4\pi e^2}{m} N(r) \, .
\end{equation}
The refraction index $n$ of this plasma is

\begin{equation} \label{plasma-n}
n^2 = 1 - \frac{\omega_e^2}{[\omega(r)]^2} \, .
\end{equation}
Here $\omega_e$ is the electron plasma frequency, $N(r)$ is the electron concentration in plasma, $e$ is the
charge of the electron, $m$ is the electron mass \footnote{In the most parts of this paper we use a convention $G=c=1$. Since the constants $G$ and $c$ are not included in the plasma frequency $\omega_e$, it looks just the same as in Gaussian cgs units. To calculate the numerical value of plasma frequency, we can just use all variables in the right hand side of (\ref{eq:omega-e}) in Gaussian cgs units. For reader's convenience, it can be used: $\omega_e \simeq 5.64 \cdot 10^4 \sqrt{N}$ sec$^{-1}$ where $N$ is measured in cm$^{-3}$ \cite{Zhelezn}. At the same time, for numerical estimates in formulas containing the black hole mass $M$ we need to restore the factors of $c$ and $G$. See, for example, formulas (68), (69), (77) in \cite{Perlick2015}.}. The frequency measured by a static observer is a function of $r$,
according to the gravitational redshift formula,

\begin{equation} \label{eq:gr-redshift}
\omega(r) = \frac{\omega_0}{\sqrt{A(r)}} \, .
\end{equation}
Here $\omega_0$ is the frequency at infinity.

Now let us restrict ourselves by the consideration of the Schwarzschild metric:

\begin{equation}
ds^2 = - A(r) \, dt^2 + \frac{dr^2}{A(r)} + r^2
\left( d \theta^2 + \sin^2 \theta d \varphi^2 \right),  \;  A(r) = 1 - \frac{2M}{r} .
\label{metric}
\end{equation}

In the case of a plasma with the refractive index
(\ref{plasma-n}), the Hamiltonian with the photon momentum 4-vector $p^i$ can be written
in the form \cite{BKTs2009, BKTs2010, Perlick2000}

\begin{equation}
H(x^i,p_i) = \frac{1}{2} \left[ g^{ik} p_i p_k + \omega_e^2 \hbar^2 \right] = 0 \, . \label{H-definition}
\end{equation}
The system of equations of motion for the space
components $x^\alpha$, $p_\alpha$ are:

\begin{equation} \label{eqs-motion-general}
\frac{dx^\alpha}{d \lambda} = g^{\alpha \beta} p_\beta \, , \quad \frac{dp_\alpha}{d \lambda}
 = -\frac{1}{2} \, g^{ik}_{,\alpha} p_i p_k - \frac{1}{2} \, \hbar^2 \left( \omega_e^2 \right)_{,
	\alpha} \, .
\end{equation}
For explicit form of these equations in Cartesian coordinate system see Refs.~\cite{BKTs2009}, \cite{BKTs2010},
in spherical coordinate system see Refs.~\cite{TsBK2013}, \cite{BKTs2015}.
The deflection angle of the photon moving from infinity to the central object and then to infinity is
\begin{equation} \label{perlick-angle}
\hat{\alpha} = 2 \int \limits_R^\infty \frac{dr}{\sqrt{r(r-2M)}\sqrt{\frac{h^2(r)}{h^2(R)} - 1 }} - \pi \, .
\end{equation}
Here we use notation
\begin{equation} \label{perlick-h}
h(r) = r \sqrt{ \frac{1}{A(r)} - \frac{\omega_e^2(r)}{\omega_0^2} } = r \sqrt{ \frac{r}{r-2M} - \frac{\omega_e^2(r)}{\omega_0^2} } \, .
\end{equation}
This expression for $\hat\alpha$ was derived for the first time in \cite{Perlick2000}, and rederived in \cite{TsBK2013} by Synge's approach.
At given $M$ and $\omega_e(r)$, the deflection angle is determined by the closest approach distance $R$ and the photon
frequency at infinity $\omega_0$. This formula allows us to calculate the deflection angle of the photon moving in the Schwarzschild metric
in presence of a spherically symmetric distribution of plasma. The expression of (\ref{perlick-angle}) via elliptic integrals for the case of homogeneous plasma can be found in \cite{TsBK2013} and \cite{BKTs2015}.

Exact expression for deflection angle in presence of Schwarzschild gravity and plasma (\ref{perlick-angle}) is a manifestation of simultaneous
and mutual action of different physical effects: gravity, refraction, dispersion. This angle (\ref{perlick-angle}) can not be calculated
analytically in a general case. For analysis and applications of this result it is convenient to consider some particular cases with appropriate
approximations. Considering the case of small total deflection angle $\hat{\alpha} \ll 1$, we have advantages to see clearly which physical
reasons give contributions to different parts of the total deflection, see Refs.~\cite{BKTs2015}, \cite{BKTs2009}, \cite{BKTs2010}.\\

\subsection{Homogeneous plasma, weak deflection} \label{subsection-hom-plasma}

In the case of a homogeneous plasma with $\omega_e^2 =$ const the refractive index does not depend on the space coordinates
explicitly, so the refractive action is absent. Therefore, in homogeneous case, we can consider the angle
$\hat{\alpha}$ as the gravitational deflection in a given medium (plasma). Despite the absence of deflection due to refraction, this deflection differs from vacuum gravitational deflection.

In the homogeneous plasma, the total deflection angle $\hat{\alpha}$ represents only the action of the gravitation. In more complicated situation, when plasma is non-homogeneous, both effects presents: difference of gravitational deflection from vacuum case due to presence of medium, and refractive deflection due to medium inhomogeneity. In weak deflection approximation it is possible to study them separately from each other.

Let us consider a situation when the closest approach distance is much larger than the Schwarzschild radius, $R \gg M$ ($R_S=2M$). It means also that $r \gg M$ (during the motion the $r$-coordinate changes from $R$ to infinity), $b \gg M$, and the resulting deflection angle is small, $\hat{\alpha} \ll 1$. Expanding the exact formula (\ref{perlick-angle}) with $\omega_e^2 =$ const, we obtain the deflection angle in the form (see Appendix A of \cite{BKTs2015}):
\begin{equation} \label{angle-plasma-R}
\hat{\alpha} = \frac{2M}{R} \left( 1 + \frac{1}{1 - \omega_e^2/\omega_0^2} \right) ,
\end{equation}
\[
\mbox{or, in ordinary units,} \; \; \hat{\alpha} = \frac{2GM}{c^2 R} \left( 1 + \frac{1}{1 - \omega_e^2/\omega_0^2} \right), \;\; R_S = \frac{2GM}{c^2}.
\]

For gravitational lensing the dependence of angles on impact parameter $b$ is needed. In approximation $R \gg M$ the difference between $R$ and $b$ is negligible, so we can just substitute $R \simeq b$ into (\ref{angle-plasma-R}), obtaining
\begin{equation} \label{angle-plasma-b}
\hat{\alpha} = \frac{2M}{b} \left( 1 + \frac{1}{1 - \omega_e^2/\omega_0^2} \right) .
\end{equation}
The deflection angle (\ref{angle-plasma-b}) can be also derived in a simpler way, see \cite{BKTs2009}, \cite{BKTs2010}. To do this, the Synge's equations have been written in Cartesian coordinates. We have considered the photon with unperturbed trajectory in a homogeneous plasma as a straight line parallel to $z$-axis with an impact parameter $b$, and have found the deflection in the approximation of small perturbations, by the integration along the straight trajectory.

Formula (\ref{angle-plasma-b}) is valid only for $\omega_0 > \omega_e$, because the waves with $\omega_0 < \omega_e$ do not propagate in the plasma. At $\omega_0$ approaching $\omega_e$, the gravitational deflection in plasma can be much larger than in the vacuum, $\hat{\alpha} \gg 2R_S/b$ (condition $\hat{\alpha} \ll 1$ should be satisfied too).
Formula (\ref{angle-plasma-b}) does not imply that $\omega_e/\omega_0 \ll 1$. If it is additionally supposed, we can rewrite the deflection (\ref{angle-plasma-b}) as:
\begin{equation} \label{angle-plasma-exp}
\hat{\alpha} = \alpha_{einst} + \alpha_{add} = \frac{2R_S}{b} + \frac{\omega_e^2}{2\omega_0^2} \frac{2R_S}{b} .
\end{equation}
Here we denote the vacuum gravitational deflection as $\alpha_{einst}$ and additional correction to the gravitational deflection connected with
the plasma presence as $\alpha_{add}$.

\emph{The presence of plasma increases the gravitational deflection angle.} In homogeneous plasma the photons of smaller frequency,
or larger wavelength, are deflected by a larger angle by the gravitating center. The effect of difference in the gravitational deflection angles
is significant for longer wavelengths, when $\omega_0$ is approaching $\omega_e$. That is possible only for the radio waves. Therefore,
the gravitational lens in plasma acts as a radiospectrometer \cite{BKTs2009}.

We should use formula (\ref{angle-plasma-b}) when we consider gravitational lensing of radiowaves by point or spherical body in presence
of homogeneous plasma. This effect has a general relativistic nature, in combination with the dispersive properties of plasma. We should
 also emphasize that the plasma is considered here like the medium with a given index of refraction, and this formula does not take
 into account gravitation of particles of plasma.
 
In the case of axisymmetric mass distribution the deflection with the
impact parameter $b$ is equal to the Einstein angle for the mass
$M(b)$, where $M(b)$ is the projected mass enclosed by the circle of
the radius $b$. In another words it is mass inside the cylinder with
the radius $b$. So, for this
case, we should write \cite{Clark1972, GL, GL2, BliokhMinakov}:
\begin{equation} \label{angle-projected-mass}
\alpha_{einst} = \frac{4 M(b)}{b} \, \; \; \mbox{or, in usual units} \; \;\frac{4 G M(b)}{c^2 b} \, .
\end{equation}

Formula (\ref{angle-plasma-b}) can be recovered with using the analogy between the motion of photon in the homogeneous plasma and massive particle in vacuum, described in the Introduction, in Subsection \ref{subsection-notes}. Test massive particle passing with the
velocity $v$ near a spherical body with the mass $M$, with the impact parameter $b \gg R_S$, is deflected by the 
angle $\alpha_m$ defined as \cite{MTW, zadachnik}:

\begin{equation}
\alpha_m = \frac{R_S}{b} \left( 1 + \frac{1}{v^2} \right) .
\end{equation}
Using group velocity of wave packet in the homogeneous plasma $v_{group} = 
[1-(\omega_e^2/\omega_0^2)]^{1/2}$, we obtain the formula (\ref{angle-plasma-b}).\\

\subsection{Non-homogeneous plasma, weak deflection} \label{subsection-nonhomog}

In the case of a plasma non-homogeneity there is also the refractive deflection, $\alpha_{refr}$.
Our approach for gravitational lensing in plasma developed in \cite{BKTs2009, BKTs2010} allows to
consider two effects simultaneously: the gravitational deflection in plasma which is different from the Einstein angle,
and the refraction connected with the plasma inhomogeneity which does not depend on the gravity. In the paper \cite{BKTs2010}
we have derived formulae for the deflection angle by a spherical body, or spherically distributed gravitating matter,
in presence of a homogeneous and non-homogeneous plasma, with taking into account the gravitation deflection
together with the refractive deflection.

In applications the usual vacuum gravitational deflection $\alpha_{einst}$ gives the major contribution
to the total deflection $\hat{\alpha}$, and all effects connected with plasma presence are significally smaller:
$\alpha_{add} \ll \alpha_{einst}$ and $\alpha_{refr} \ll \alpha_{einst}$.

Additional correction $\alpha_{add}$ to the gravitational deflection caused by plasma is in most cases smaller
than correction due to refraction $\alpha_{refr}$ (note that both corrections have the same frequency dependence, they are proportional to $1/\omega_0^2$). In these cases the total deflection angle (in a weak deflection
approximation) can be written as a sum of vacuum gravitational deflection $\alpha_{einst}$ and refractive deflection $\alpha_{refr}$.
In this approximation  $\alpha_{add}$ is neglected, and the gravitational deflection is taken as equal to vacuum gravitational
deflection $\alpha_{einst}$ \cite{BliokhMinakov}.

Expanding the exact integral formula (\ref{perlick-angle}) with $M/r \ll 1$ and $\omega_e^2(r)/\omega_0^2 \ll 1$, we obtain an approximate formula for the case of a weak deflection in non-homogeneous plasma (see Appendix B in \cite{BKTs2015}):
\begin{equation} \label{angle-nonhomogen-r}
\hat{\alpha} =  \alpha_{einst} + \alpha_{refr} \, , \;
\alpha_{einst} = \frac{4M}{R} = \frac{2R_S}{R} \, , \; \alpha_{refr} =  \frac{R K_e}{\omega_0^2} \int \limits_R^\infty \frac{1}{\sqrt{r^2-R^2}} \frac{dN(r)}{dr} \, dr \, , \; K_e \equiv \frac{4 \pi e^2}{m} \, .
\end{equation}

For gravitational lensing the dependence of angles on the impact parameter $b$ is needed. In papers \cite{BKTs2009} and \cite{BKTs2010} we have derive $\hat{\alpha}$ using Cartesian coordinates. We have considered the photon with the unperturbed trajectory as a straight line parallel to $z$-axis, with the impact parameter $b \simeq R \gg M$. In this case we have, at given $b$:

\begin{equation} \label{angle-nonhomogen-b}
\hat{\alpha} =  \alpha_{einst} + \alpha_{refr} \, , \quad
\alpha_{einst} = \frac{4M}{b} = \frac{2R_S}{b} \, , \quad \alpha_{refr} =  \frac{K_e}{\omega_0^2}   \int \limits_0^\infty \frac{\partial N}{\partial b} \, dz .
\end{equation}
Here $N=N(r)$ and $r=\sqrt{b^2+z^2}$, so the expression under the integral sign is a function of $b$ and $z$. To calculate the deflection angle, 
we perform partial differentiation with respect to $b$ at constant $z$ and then, after integration along $z$ at constant $b$, 
obtain deflection angle as a function of $b$. In Appendix B of \cite{BKTs2015} we have also shown how to transform $\alpha_{refr}$ in the form (\ref{angle-nonhomogen-r}) into form (\ref{angle-nonhomogen-b}).

The presence of homogeneous or non-homogeneous plasma increases the gravitational deflection of photons. 
A vacuum gravitational deflection is usually considered as positive ($\alpha_{einst}>0$) therefore the additional correction to 
gravitational deflection due to plasma presence is also positive ($\alpha_{add}>0$).
The refractive deflection can be both positive or negative, depending on the density profile. Usually the density of plasma in 
different models decreases with radius ($dN/dr<0$), therefore the refractive deflection is usually opposite to the gravitational 
deflection: the correction due to refractive deflection is negative ($\alpha_{refr}<0$), see \cite{BKTs2010}.

For example, in the case of inhomogeneous plasma with a power-law concentration
\begin{equation}
N(r) = N_0 \left( \frac{r_0}{r} \right)^k, \;  N_0 = \mbox{const}, \; r_0 = \mbox{const}, \; k=\mbox{const}>0
\label{k}
\end{equation}
the refractive deflection is
\begin{equation} \label{angle-refr-power}
\alpha_{refr} = - \frac{K_e}{\omega_0^2} N_0 \left(\frac{R_0}{b}\right)^k \frac{\sqrt{\pi} \,
	\Gamma\left(\frac{k}{2} + \frac{1}{2}\right)}{\Gamma\left(\frac{k}{2}\right)} \, , \quad \Gamma(x) = \int \limits_0^\infty t^{x-1} e^{-t} dt .
\end{equation}\\


\section{Effects of plasma in strong lens system}

\subsection{Shift of angular position of image}

\begin{figure}[h]
	\centering
	\includegraphics[width=14cm]{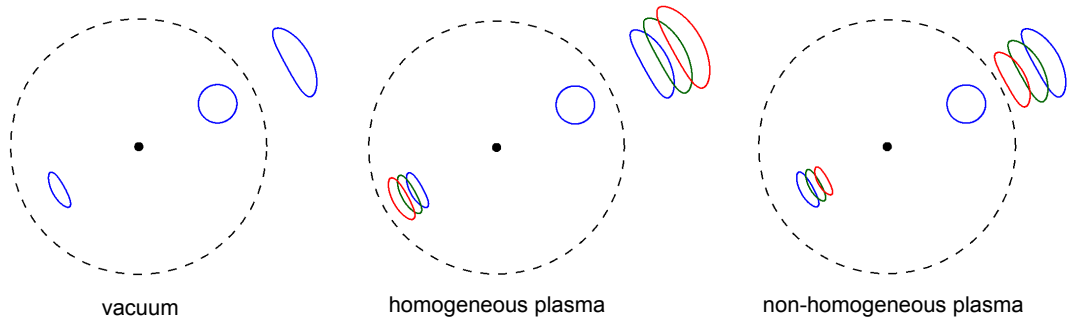}
	\caption{Comparison of lensing by point mass in vacuum (left) and in plasma (middle and right). The photon trajectories and deflection 
angles in vacuum don't depend on the photon frequency (or energy). Therefore lensing in vacuum is achromatic, and the Schwarzschild 
point-mass lens leads to two images with complicated spectra, formed by the photons of different frequencies which undergo deflection
 by the same angle. If lens is surrounded by plasma, the light rays of different frequencies have different deflection angles. The 
 images become diffused and divide into many images formed by photons of different frequencies. Difference in angular position of images of different frequencies take place both in homogeneous and non-homogeneous plasma and is a subject for observation. On the left figure we present the vacuum lensing: a blue circle is the source, a black spot is the lens, a dashed circle is the Einstein ring. In the middle we present schematic picture for the situation when lens is surrounded by homogeneous plasma, see subsection \ref{subsection-hom-plasma}. Now blue color is used for images with formally infinite photon frequency (for optical frequencies the effect of plasma presence is negligible so for these frequencies it corresponds to vacuum case). Photons of lower frequencies (green and red colors) undergo plasma influence and their deflection angles becomes larger. Therefore green and red images are further from the center of the picture. Angular separation between two images of the same color becomes larger with decrease of the wave frequency (or increase of the wavelength). On the right picture we present the situation when the lens is surrounded by non-homogeneous plasma, and refraction effects prevail over effects of additional gravitational deflection due to plasma, see subsection \ref{subsection-nonhomog}. Now plasma presence leads to decrease of the total deflection, therefore green and red images are located closer to the center of the picture. Angular separation between two images of the same color becomes smaller with decrease of the photon frequency. This case might be observed by modern facilities. Note that in these pictures we assume that the number and structure of the images do not change drastically due to the presence of plasma. In these pictures we also assume that the emission regions of the source in different bands coincide, which may not necessarily be the case in observations.} 
 \label{figure-spectr}
\end{figure}

Presence of plasma always makes gravitational lensing chromatic. In the case of a homogeneous plasma the chromatic effects arise due to 
difference of gravitational deflection itself from vacuum case. In non-homogeneous plasma case the chromatic refractive deflection takes 
place in addition. The observational effect of the frequency dependence may be explained on the example of the Schwarzschild point-mass 
lens \cite{BKTs2009, BKTs2010}. Instead of two concentrated images with complicated spectra, we will have two expanded images, formed by 
the photons with different frequencies, which are deflected by different angles (Fig.\ref{figure-spectr}). To see the difference in 
angular position of images, we should compare radio and optical observations of images. For optical frequencies the effect of plasma 
presence is negligible, and positions of images in this case can be calculated with the vacuum formulae.

In the case of strong lens system, it is sufficient to use the weak deflection approximation. For homogeneous plasma and point mass lens 
the formula (\ref{angle-plasma-b}) can be used. If it is additionally assumed that plasma frequency is much smaller than the photon 
frequency, the formula (\ref{angle-plasma-exp}) can be used. For non-homogeneous plasma and point mass lens the formula 
(\ref{angle-nonhomogen-b}) is sufficient. For power-behaved distribution of plasma the refractive deflection is calculated 
by (\ref{angle-refr-power}). If gravitating mass is distributed, for calculation of the gravitational deflection see 
(\ref{angle-projected-mass}).

For observational prediction of plasma effects described above we refer to work of Er and Mao \cite{Mao2014}
 who have performed the numerical modelling of plasma effects in strong lens system similar to Einstein Cross. 
 Authors have used realistic distribution of plasma for spiral and elliptical galaxies, and have investigated numerically 
 the influence of plasma presence on position of the images.
The total deflection angle was calculated as a sum of vacuum gravitational deflection and refractive plasma deflection, with using formulas 
from \cite{BKTs2009}, \cite{BKTs2010}. For calculation of gravitational deflection the lens at redshift 0.5, and the source at redshift 2.0
 were considered, with a size of the vacuum Einstein ring equal to 0.4 arcsec.

For refractive plasma deflection authors have considered two plasma models. The first one is for a spiral lensing galaxy 
with the plasma concentration
\begin{equation} \label{Mao-1}
N(r) = N_0 \, e^{-r/r_0} ,
\end{equation}
where $r$ is the radial distance from the center of the galaxy, and constants are $N_0 = 10$ cm$^{-3}$, and $r_0 = 10$ kpc. 
The second plasma model is used as appropriate for elliptical galaxies:
\begin{equation} \label{Mao-2}
N(r) = N_0 \, (r/r_0)^{-1.25},
\end{equation}
with $N_0 = 0.1$ cm$^{-3}$ and $r_0 = 10$ kpc. These values of $N_0$ correspond to the plasma frequency $\omega_e \simeq 1.8 \cdot 10^5$ sec$^{-1}$ and $1.8 \cdot 10^4$ sec$^{-1}$.
Authors have found that for the radio frequency 375 MHz of the photon the difference in image positions due to plasma presence can be a few 
tenth milliarcsecs (for the first model), what is possible to detect for the low frequency radio observations \cite{Mao2014}. Similar changes will happen 
for the magnification of the lensed images, but there are also some other factors which can change the magnification more significantly, 
such as the substructures in the lens galaxy or cluster.\\

To conclude this Section, let us formulate the following way of observation of plasma effects in strong lens systems:

(i) Compare observations of strong lens system with
multiple images in optical and radio band, or compare observations in two radio bands;

(ii) Shift of the angular position of every image can be observed.

\noindent As a result, we could make estimations of plasma properties in the vicinity of lens. Note that the emission regions of the source in different bands may not coincide. Speaking about quasars we can expect that the centers of emission regions will be coincide due to symmetry of accretion disc, and only jets separated from the central black hole can have angular position different from the main source. Nevertheless, even in this case plasma effects can be extracted, assuming spherically symmetric distribution of plasma. Its presence leads to a shift of every image mainly in the plane passing through observer, lens and this image, whereas the change of the source position leads to more complicated changes of image positions.
 \\

\subsection{Effect on flux ratio in different wavelengths}

Gravitational lensing also leads to magnification of the source. This means that the flux of the image is larger or smaller than the flux 
of the unlensed source, and the ratio of these fluxes is referred as magnification. Magnification is determined by the ratio of solid angles of lensed image and unlensed source, different images have different magnifications.
The magnification properties are determined by the deflection law, and the magnification of every image can be calculated for given lens system.

In observations we don't know the intrinsic flux of the source and its spectrum, therefore we can not observe the magnification of every image directly. What we can do is to observe and compare the fluxes of images; ratio of fluxes of different images equals to the ratio of calculated magnifications of these images.

For study of plasma effects we need to investigate the flux ratio in different bands. In vacuum the gravitational deflection is achromatic. Therefore the ratios of the fluxes of images in different bands should be the same, 
if we consider lensing in vacuum.

For example, let us consider two observed images, 1 and 2, in strong lens system. Fluxes $F$ of images in optical and radio frequences will have the same ratios:
\begin{equation}
\frac{F_1^{opt}}{F_2^{opt}} = \frac{F_1^{rad}}{F_2^{rad}} \, .
\end{equation}

This situation is changed if we assume that the lens is surrounded by plasma. In the case of lensing of radio waves in the plasma we have another formula for 
deflection instead of the Einstein angle, so formulas for magnification will be also different from vacuum formulas. 
In general case, the light rays from every image propagate through plasma regions of different density. Magnification of every image is therefore changed differently in different bands (for example, for given observational frequency one image can be strongly affected by the plasma presence but for another image the effect can be negligible). By-turn it leads to difference 
of ratios of the fluxes of images in different bands (see details in \cite{BKTs2010}). Note that the flux ratios give an information on the plasma concentration only in the vicinity of images (light rays from different images propagate through different plasma regions), not the whole lens. For simplicity, for estimates we can use spherically symmetric distribution of plasma, see (\ref{Mao-1}) and (\ref{Mao-2}).

For example, for two images in strong lens system we will have:
\begin{equation} \label{ineq-magn}
\frac{F_1^{opt}}{F_2^{opt}} \neq \frac{F_1^{rad}}{F_2^{rad}} \, .
\end{equation}

Of course, there exist other reasons of arising this inequality. First of all, Thompson scattering and absorption during propagation of radiation through the plasma can significantly (and chromatically) change the flux and complicate the phenomena of lensing magnification.

The second thing is sometimes referred as 'chromatic microlensing' of quasars. 'Macrolensing' by galaxy (lens) leads to several images of the quasar, but every image is affected by 'microlensing' by individual stars in lens. Microlensing influences the spectrum and the brightness of the lensed image. Physical sources like quasars may have different sizes for different wavelengths (chromatic profile). Magnification and the light curves expected as a result of microlensing depend on the source size, so chromatic effects arise. Such chromatic effects can give information about source structure, see \cite{Kayser1986}, \cite{WambsganssPaczynski1991}.

To conclude, if someone observe the inequality (\ref{ineq-magn}) in strong lens system, there may be the following possible explanations:

(i) extinction (absorption and scattering) which can be chromatic;

(ii) microlensing of lensed image by individual stars in lens which can be chromatic due to physical properties of quasar; 

(iii) chromatic gravitational lensing due to plasma presence in vicinity of lens.\\


\section{Lensing by black holes in presence of plasma}

Let us consider the Schwarzschild black hole and discuss an influence of plasma on positions and magnifications of high-order (relativistic) images. In the case of a weak deflection of the photon, the deflection angle is calculated by the Einstein formula. In Darwin's paper \cite{Darwin1959} 
it was shown that another limiting situation,  for the
photons which undergo  several circles around the central object, can also be investigated analytically.
In this limit, which is called strong deflection limit, the deflection angle is written as \, \cite{Darwin1959, Bozza2001}
\begin{eqnarray}
\label{vacuum-alpha-R}
\hat{\alpha} &=& - 2 \ln \left( \frac{R}{r_M} - 1\right) +  2 \ln[12(2-\sqrt{3})] - \pi = \nonumber\\
&=& - 2 \ln \frac{R-3M}{36(2-\sqrt{3}) M} - \pi = - 2 \ln \frac{(R-3M)(2+\sqrt{3})}{36M} - \pi \, ,
\end{eqnarray}
or, as a function of the impact parameter $b$ \, \cite{Bozza2001,Bozza2002}
\begin{eqnarray}
\label{vacuum-alpha-b}  \hat{\alpha} &=& - \ln \left( \frac{b}{b_{cr}} - 1 \right) + \ln[216(7-4\sqrt{3})] - \pi \, = \nonumber\\
&=& - \ln \frac{b-3\sqrt{3}M}{648\sqrt{3}(7-4\sqrt{3})M} - \pi   = - \ln \frac{(b-3\sqrt{3}M)(7+4\sqrt{3})}{648\sqrt{3}M} - \pi \, .
\end{eqnarray}

\begin{figure}[h]
	\centering
	\includegraphics[width=12cm]{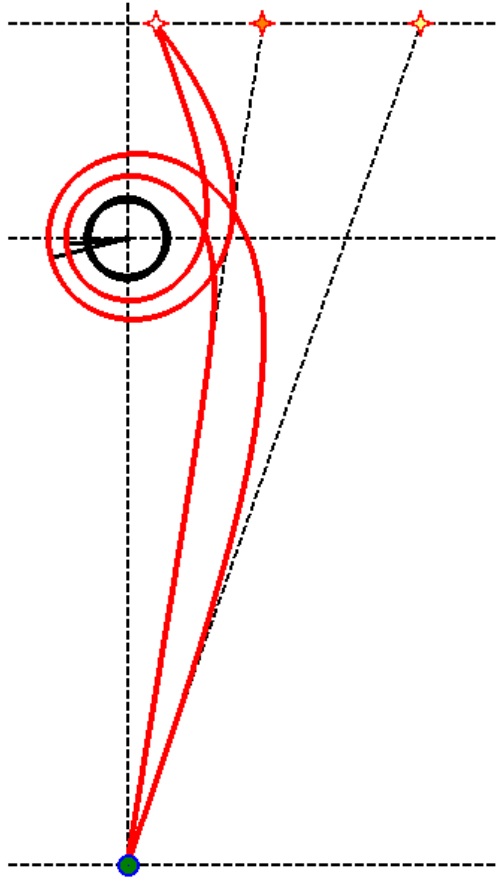}
	\caption{Relativistic images in the case of lensing by black hole surrounded by homogeneous plasma. 
Two images and corresponding light rays of different wavelengths are calculated.} \label{figure-rel-plasma}
\end{figure}

Let us discuss now the influence of plasma presence on relativistic images (Fig. \ref{figure-rel-plasma}).
The photon deflection angle in homogeneous plasma in the strong deflection limit as
a function of the closest approach distance $R$, and the ratio of frequencies
$\omega_e^2/\omega_0^2$ has a form \cite{TsBK2013}:

\begin{equation} \label{alpha-R-E}
\hat{\alpha}(R,x) = -2 \sqrt{\frac{1+x}{2x}} \ln \left[z_1(x) \, \frac{R-r_M}{r_M}  \right] - \pi \, ,
\end{equation}
where

\begin{equation} \label{z-x}
\quad z_1(x) = \frac{9x-1+2\sqrt{6x(3x-1)}}{48x} \, , \quad
r_M=6M \, \frac{1+x}{1+3x} \, , \quad x \equiv \sqrt{1-\frac{8 \omega_e^2}{9 \omega_0^2}} \, .
\end{equation}
The formula (\ref{alpha-R-E}) is asymptotically exact, and is valid for $R$ close to $r_M$. Critical (minimum) 
value of $R$ is equal to $r_{M}$ which is a radius of the point where a maximum of the effective potential takes place. 
The deflection angle of a photon goes to infinity when $R$ goes to $r_M$, and photon performs infinite number of turns 
at the radius $r_{M}$. If $\omega_0 \gg \omega_e$, then $x \to 1$, giving $r_{M} = 3M$, what corresponds to the photon in the vacuum.

The deflection angle in strong deflection limit as a function of the impact parameter $b$ and the ratio of frequencies is
$\omega_e^2/\omega_0^2$ is written as

\begin{equation} \label{alpha-b-E}
\hat{\alpha}(b,x) = - \sqrt{\frac{1+x}{2x}} \, \ln \left[ \frac{2\,z_1^2(x)}{3x} \, \frac{b-b_{cr}}{b_{cr}} \right] - \pi \, , \quad  \mbox{where} \;\;  b_{cr} = \sqrt{3} \, r_M \, \sqrt{\frac{1+x}{3x-1}} \, .
\end{equation}
This formula is valid for $b$ close to $b_{cr}$, where $b_{cr}$ is a critical value of impact parameter under given $\omega_e^2/\omega_0^2$. For $\omega_0 \gg \omega_e$ we obtain the critical
impact parameter for vacuum, $b_{cr} = 3 \sqrt{3} M$.

Formula (\ref{alpha-b-E}) can be applied for calculation of positions of relativistic images in presence of a homogeneous
plasma. For simplicity, let us rewrite formula (\ref{alpha-b-E}) as

\begin{equation} \label{alpha-simple}
\hat{\alpha}(b,x) = - a(x) \, \ln \left( \frac{b-b_{cr}}{b_{cr}} \right) + c(x) \, ,
\end{equation}
where $a(x)$ and $c(x)$ are defined as

\begin{equation}
a(x) = \sqrt{\frac{1+x}{2x}} \, , \quad c(x) = - \sqrt{\frac{1+x}{2x}} \, \ln \left[ \frac{2\,z_1^2(x)}{3x} \right] - \pi \, .
\end{equation}
Solving the equation

\begin{equation} \label{2-pi-n}
\hat{\alpha}(b,x) = 2 \pi k \, , \quad k=1,2, ... \, ,
\end{equation}
where $k$ is a number of pairs of relativistic images, we obtain the expressions for 
impact parameters $b_k(x)$, and the corresponding angular positions of the relativistic images $\theta_k(x)$,
in the form

\begin{equation}
b_k(x) = b_{cr} \left[ 1 + \exp \left(\frac{c(x)-2\pi k}{a(x)} \right) \right] \,  , \quad
\theta_k(x) = \frac{b_{cr}}{D_d} \left[ 1 + \exp \left(\frac{c(x)-2\pi k}{a(x)} \right) \right]  \,  ,
\label{theta-n}
\end{equation}
Here $D_d$ is the distance between the observer and the lens. The angular positions $\theta_k$ in plasma are 
always bigger than the positions in vacuum. Therefore the presence of homogeneous plasma \textit{increases} the 
angular separation of the point relativistic images from gravitating center or the angular size of the relativistic rings.

\begin{figure}
	\centering
	\includegraphics[width=13cm]{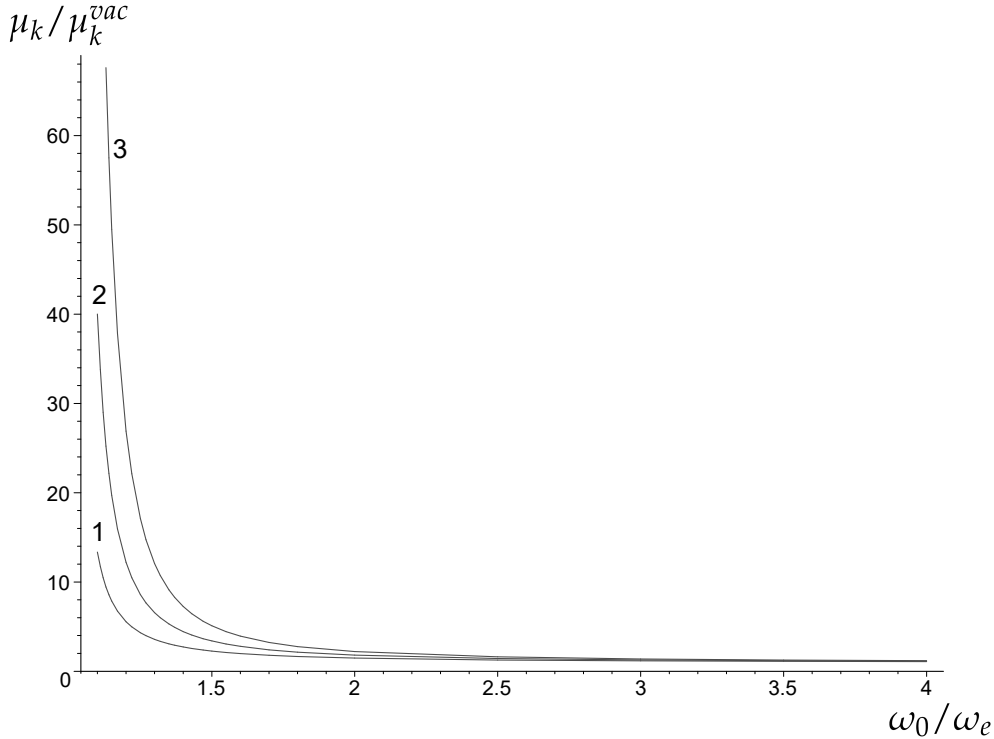}
	\caption{Comparison of the magnification factors of relativistic images for lensing in homogeneous plasma and in vacuum. Ratios $\mu_k/\mu_k^{vac}$ are plotted as a function of $\omega_0/\omega_e$, for $k=1,2,3$. Here $\mu_k^{vac}$ is the vacuum value of magnification. Values $\mu_k$ and $\mu_k^{vac}$ depend on $D_s$, $D_d$, $D_{ds}$, $\beta$, $\omega_0/\omega_e$, but ratio $\mu_k/\mu_k^{vac}$ depends only on $\omega_0/\omega_e$. }
	\label{figure-magnification}
\end{figure}

The magnification of the relativistic images of a point source located at the angular position $\beta$ from 
line connecting the observer and the gravitational center (lens) is equal to \cite{TsBK2013}

\begin{equation} \label{mu-n}
\mu_k = \frac{D_s b_{cr}^2 (1+ l_k) l_k}{D_{ds} D_d^2  \, a(x) \, \beta} \, , \quad    \mbox{with} \;\; l_k = \exp \left(\frac{c(x)-2\pi k}{a(x)} \right) ,
\end{equation}
and $a(x)$ and $c(x)$ are coefficients defined in (\ref{alpha-simple}). In the expression (\ref{mu-n}) the variables 
$b_{cr}$, $l_k$, $a(x)$, $c(x)$ depend on $x$, so these variables depend on the ratio of the photon and the plasma frequencies. 
Magnification $\mu_k$ of the relativistic images tends to infinity if the source angular position $\beta$ goes to 0, as it takes place 
in the vacuum \cite{Virbhadra2000}. If $\omega_0 \rightarrow \omega_e$, magnifications $\mu_k$ increases unboundedly, because $b_k$ tends to 
infinity in this limit, according to (\ref{z-x}), (\ref{alpha-b-E}), see Fig. \ref{figure-magnification}. Although presence of plasma can significantly increase the magnifications in comparison with plasma, the magnification remains very small because vacuum values are very small. For example, for $M/D_d = 2.26467 \cdot 10^{-11}$, what corresponds to supermassive black hole in center of Milky Way, $D_s/D_{ds}=2$, $\beta = 1 \; \mu$as (these parameters have been taken from \cite{Virbhadra2009}) the vacuum values of magnification are \cite{TsBK2013}:
\begin{equation} 
\mu_1^{vac} =  0.716 \cdot 10^{-11}, \quad \mu_2^{vac} =  0.134 \cdot 10^{-13}, \quad \mu_3^{vac} = 0.249 \cdot 10^{-16} .
\end{equation}

\section{Influence of a plasma on the shadow of a\\ spherically symmetric black hole}
\label{section-shadow-schw}

\subsection{General formulas for spherically symmetric black holes}

\begin{figure}[h]
	\centering
	\includegraphics[width=14cm]{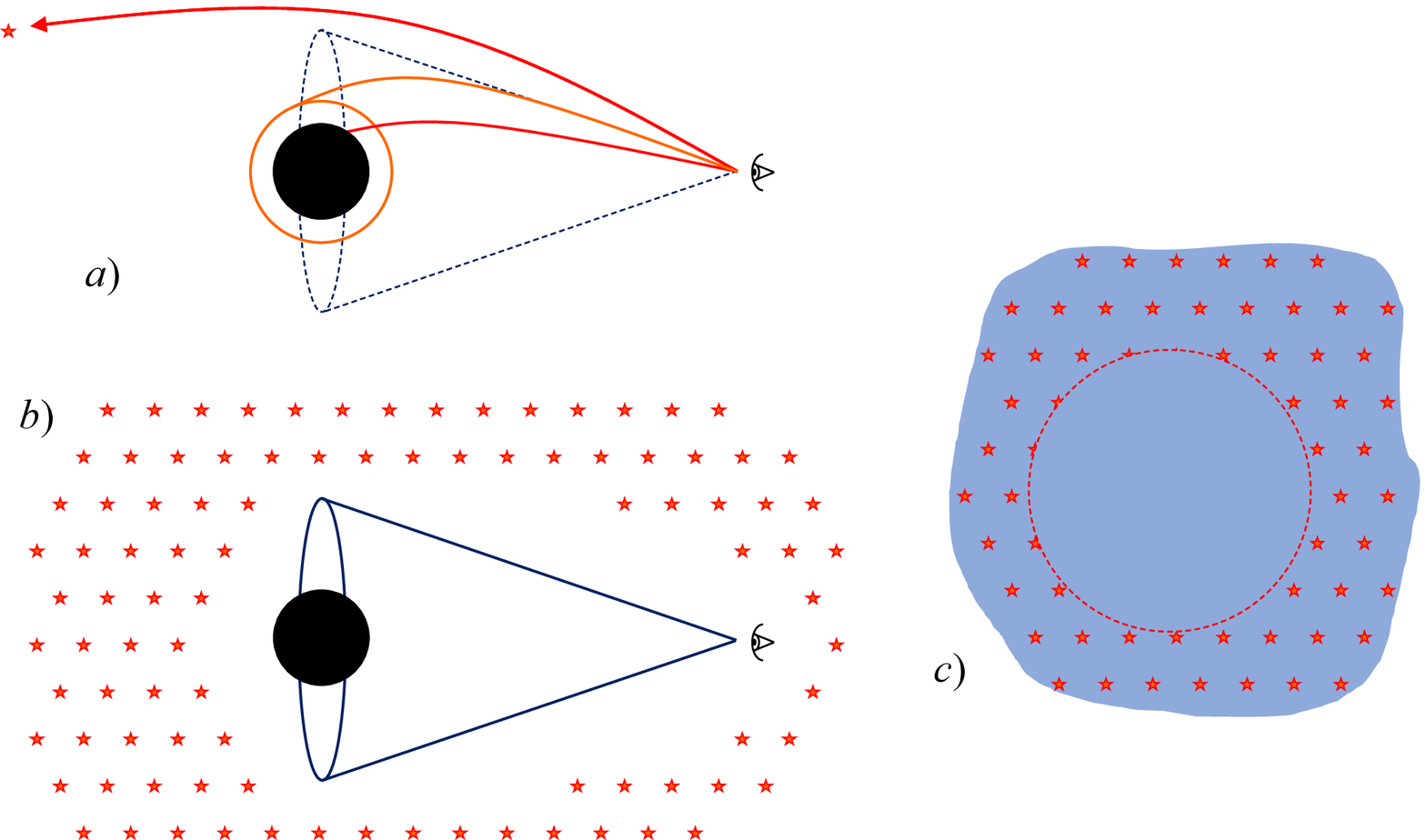}
	\caption{Formation of the black hole shadow. The shadow is the black disk an observer sees in the sky if a black hole is viewed against a backdrop of light sources that are distributed around the black hole but not between the observer and the black hole. (\textbf{a}) Let us consider light rays sent from an observer with chosen position into the past. These light rays can be divided into two classes: Light rays of the first class go to infinity after being deflected by the black hole. Light rays of the second class go towards the horizon of the black hole. The boundary between these two classes is determined by the rays that asymptotically spiral towards the outermost photon sphere. (\textbf{b}) Now let us consider that there are many light sources distributed everywhere around the black hole. We assume that there are no light sources between the observer and the black hole (speaking more strictly, there are no light sources in the region filled by the above mentioned light rays of the second class). Initial directions of rays of the second class correspond to darkness on the observer’s sky. Therefore, the cone which was fulfilled by the rays of the second class will be dark for the observer. (\textbf{c}) The observer will see the dark disc in the sky against the backdrop of light sources. Note that the picture is drawn as if the stars represent a continuous distribution of light, therefore stars positions are not affected by lensing on this picture. }
\label{figure-shadow-formation}	
\end{figure}

\begin{figure}[h]
	\centering
	\includegraphics[width=14cm]{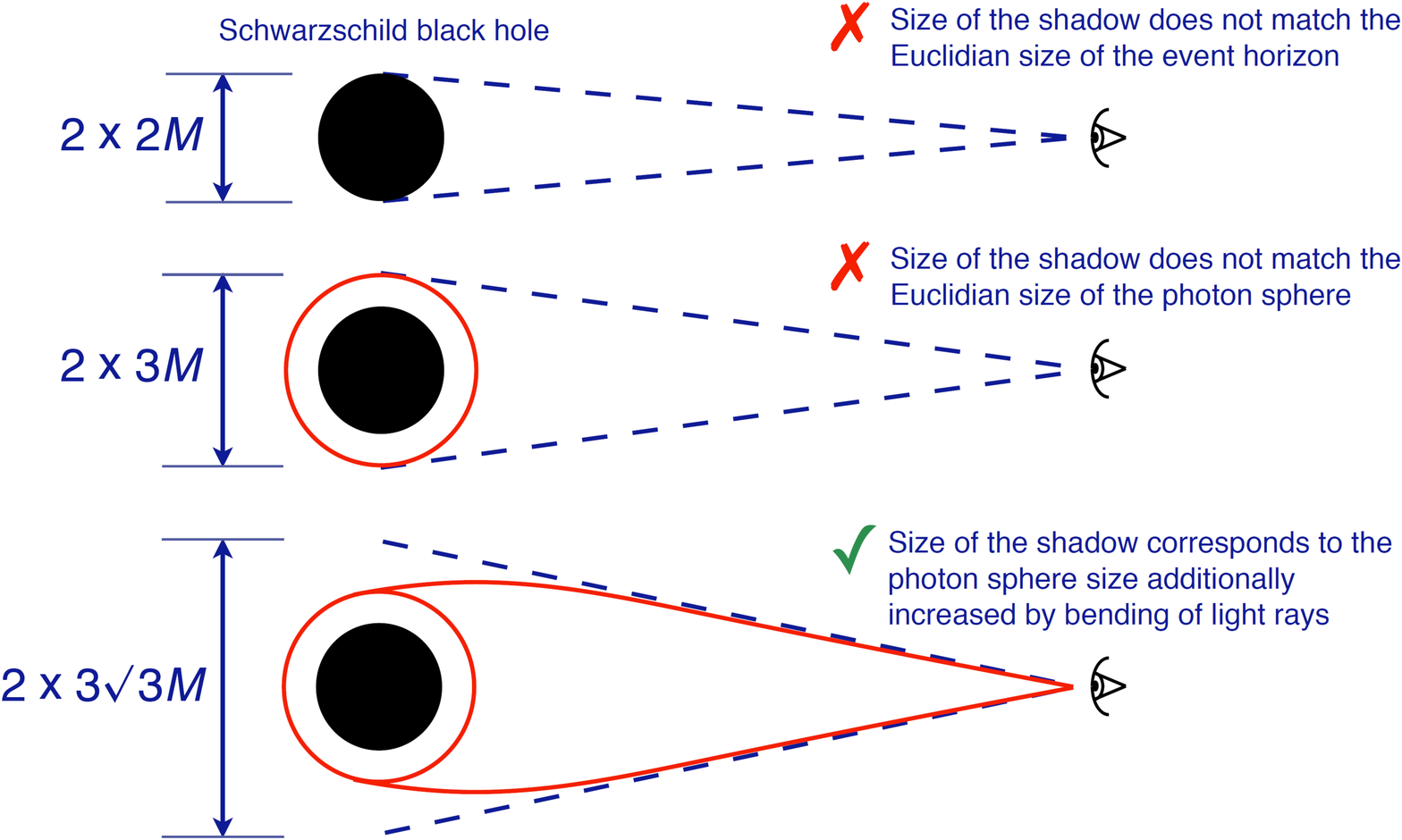}
	\caption{The shadow is bigger than just an size of black hole horizon.}
	\label{figure-not-shadows}	
\end{figure}

The shadow is defined as the region of the observer’s sky that is left dark if there are light sources distributed everywhere 
but not between the observer and the black hole. For constructing the shadow we have to consider all past-oriented light rays 
that issue from a chosen observer position. Each of these light rays corresponds to a point on the observer's sky. We assign 
darkness to a point if the corresponding light ray goes to the horizon of the black hole, and brightness otherwise. Observer
 will see a dark spot (which we call as shadow) in the angular direction where the black hole is located. The boundary of the 
 shadow is determined by the initial directions of light rays that asymptotically spiral towards the outermost photon sphere 
 (see Fig. \ref{figure-shadow-formation} and \ref{figure-not-shadows}).

The first steps towards an investigation of the shadow in
matter based on analytical calculations have been done in the paper of Perlick, Tsupko and Bisnovatyi-Kogan \cite{Perlick2015}. In that paper
we have considered the simplest non-trivial case: the influence of a non-magnetized pressure-less
plasma on the size of the shadow of a non-rotating black hole was analytically calculated.

Let us consider a spherically symmetric and static metric (\ref{eq:g}) with the refraction index (\ref{plasma-n}).
Let us define for convenience a function $h(r)$ which contains all information about spacetime and plasma and is given by
\begin{equation} \label{eq:h-definition}
h^2(r) = \frac{D(r)}{A(r)} \left( 1 - A(r)
\frac{\omega _e (r)^2}{\omega^2 _0} \right) \, .
\end{equation}
This is more general definition of function $h(r)$ given by (\ref{perlick-h}).

We obtain that the angular radius $\alpha_{\mathrm{sh}}$ of the shadow is determined by compact formula \cite{Perlick2015}:
\begin{equation}\label{eq:shadow}
\mathrm{sin} ^2 \alpha_{\mathrm{sh}} \, = \,
\frac{h^2(r_{\mathrm{ph}})}{h^2(r_{\mathrm{O}})}  \, .
\end{equation}
Here $r_{\mathrm{O}}$ is the observer position, and $r_{\mathrm{ph}}$ is the radius of photon circular orbit (for given space-time and plasma distribution) and can be found from equation
\begin{equation}\label{eq:circ8}
0 \, = \, \frac{d}{dr} h^2(r) \, .
\end{equation}
Any solution $r=r_{\mathrm{ph}}$ of (\ref{eq:circ8})
determines the radius of a \emph{photon sphere}. If a
light ray starts tangentially to such a sphere it will stay
on a circular path with radius $r_{\mathrm{ph}}$ forever.

To find
$\alpha_{\mathrm{sh}}$ for a given metric of the form (\ref{eq:g}),
a given plasma concentration $N(r)$, a given photon frequency
at infinity $\omega_0$ and a given observer position $r_{\mathrm{O}}$,
we have to calculate $r_{\mathrm{ph}}$ using eq.(\ref{eq:circ8}) and
to substitute the result into formula (\ref{eq:shadow}). Note that
the photon frequency at the observer position is $\omega(r_{\mathrm{O}})$
according to (\ref{eq:gr-redshift}).\\

\subsection{Schwarzschild black hole}

For the Schwarzschild spacetime,
\begin{equation}\label{eq:Ss}
A(r) = B(r)^{-1} = 1- \frac{2M}{r} \, , \quad
D(r) = r^2 \, ,
\end{equation}
the function $h(r)$ specifies to
\begin{equation}\label{eq:Ssh}
h^2(r) = r^2 \left( \frac{r}{r-2M} - \frac{\omega^2 _e (r)}{\omega^2 _0} \right) \, .
\end{equation}
The angular radius of the shadow (\ref{eq:shadow}) specifies to
\begin{equation}\label{eq:Schwsh}
\mathrm{sin} ^2 \alpha _{\mathrm{sh}} =
\frac{
r_{\mathrm{ph}}^2
\left( \frac{r_{\mathrm{ph}}}{r_{\mathrm{ph}}-2M} -
\frac{\omega^2 _e (r_{\mathrm{ph}})}{\omega _0^2} \right)
}{
r_{\mathrm{O}}^2
\left( \frac{r_{\mathrm{O}}}{r_{\mathrm{O}}-2M} -
\frac{\omega^2 _e (r_{\mathrm{O}})}{\omega _0^2} \right)
}
\end{equation}
where $r_{\mathrm{ph}}$ has to be determined from
(\ref{eq:circ8}) which is simplified to
\begin{equation}\label{eq:circSchw}
0 = \frac{r \, (r-3M)}{(r-2M)^2}
- \, \frac{\omega^2_e (r)}{\omega _0 ^2}
- \, r \, \frac{\omega _e (r) \omega _e ' (r)}{\omega _0 ^2} \, .
\end{equation}
For vacuum, $\omega _e (r) = 0$, our consideration
gives
\begin{equation}\label{eq:Sssh}
h^2(r) = \frac{r^2}{1 - 2M/r} \, , \quad    r_{\mathrm{ph}}=3M \, , \quad    \mathrm{sin} ^2 \alpha_{\mathrm{sh}} \, = \,
\frac{27M^2 (1-2M/r_{\mathrm{O}})}{r_{\mathrm{O}}^2} \, .
\end{equation}
This is Synge's \cite{Synge1966} formula for the radius of the shadow
of a Schwarzschild black hole which was mentioned already in the
introduction.

It is shown in paper \cite{PerlickTsupko2017} that the plasma has an increasing effect of the shadow if (inequality is correct for small plasma density case)
\begin{equation}\label{eq:increase}
\omega^2 _e \big( r_{\mathrm{O}}  \big)
\Big( 1- \frac{2 \, m}{r_{\mathrm{O}}} \Big) >
\omega^2 _e \big( r_{\mathrm{ph}}  \big)
\Big( 1- \frac{2 \, m}{r_{\mathrm{ph}}} \Big) \, .
\end{equation}
Particular cases:

(i) In case of homogeneous plasma $\omega_e = $ const, and condition (\ref{eq:increase}) is always satisfied 
as $r_{\mathrm{O}}> r_{\mathrm{ph}}$. Therefore presence of homogeneous plasma makes the shadow bigger.

(ii) In case of non-homogeneous plasma with given density distribution both increasing and decreasing effects are possible, 
depending on the position of the observer. Near the black hole, relativistic effects predominate. As we know, gravitational 
deflection becomes larger in presence of (homogeneous or non-homogeneous) plasma, therefore the shadow becomes larger for 
the observer who is close to the black hole. Far from the black hole, refraction effects predominate. Refraction on the 
declining density profile reduces the total deflection angle. Therefore, the shadow becomes smaller for the observer who 
is far from the black hole. These conclusions are evident from the formula (\ref{eq:increase}). The factor in brackets 
on the left-hand side of the inequality is always greater than the factor in brackets on the right-hand side of the inequality, 
but it does not exceed unity. At the same time, the plasma frequency on the left side of the inequality decreases as the 
observer moves away from the black hole. This leads to the fact that when the observer is moving further and further from 
the black hole, the shadow first increases, and after reaching a certain distance (which can be found from equating the 
left and right parts) decreases. Dependence of this distance on $k$ for power-law distribution (\ref{k}) is presented in
the paper \cite{PerlickTsupko2017}.\\

\subsection{Small density plasma case}

Special attention was given to the realistic case when the plasma
frequency is much smaller than the photon frequency.
If the plasma frequency is much smaller than the
photon frequency, the equations for the photon
sphere and for the radius of the shadow can be linearized around
the corresponding values for vacuum light rays. As an example, we consider the Schwarzschild
spacetime for the case that the plasma electron
density is given by a power law,

\begin{equation}
\frac{\omega _e (r) ^2}{\omega _0 ^2}  = \beta _0 \frac{M^k}{r^k}
\end{equation}
where $\beta _0 > 0$ and $k \ge 0$ are dimensionless constants. The first-order equation for the radius of the shadow yields \cite{Perlick2015}

\begin{equation}\label{eq:shadow-power}
\mathrm{sin} ^2 \alpha _{\mathrm{sh}} = \frac{27 M^2}{r_{\mathrm{O}}^2} \Big( 1 - \frac{2M}{r_{\mathrm{O}}} \Big)
\left( 1- \frac{\beta _0}{3^{k+1}}
+ \Big( 1 - \frac{2M}{r_{\mathrm{O}}} \Big)
\frac{\beta _0 M^k}{r_{\mathrm{O}}^k}
\right) \, .
\end{equation}
If the observer is far away from the black hole, $r_{\mathrm{O}} \gg M$,
this can be simplified to

\begin{equation}\label{eq:shadow-distant-hom}
\mbox{homogeneous plasma}, \quad k=0: \quad \mathrm{sin} ^2 \alpha _{\mathrm{sh}} =
\frac{27 M^2}{r_{\mathrm{O}}^2}
\left( 1 +  \frac{2 \beta _0}{3} \right) \, ,
\end{equation}

\begin{equation}\label{eq:shadow-distant-non-hom}
\mbox{non-homogeneous plasma}, \quad k>0: \quad \mathrm{sin} ^2 \alpha _{\mathrm{sh}} =
\frac{27 M^2}{r_{\mathrm{O}}^2}
\left( 1- \frac{\beta _0}{3^{k+1}} \right) \, .
\end{equation}
Again, we see that the shadow for distant observer becomes bigger in homogeneous plasma and becomes smaller in non-homogeneous one.

\begin{figure}
	\centering
	\includegraphics[width=14cm]{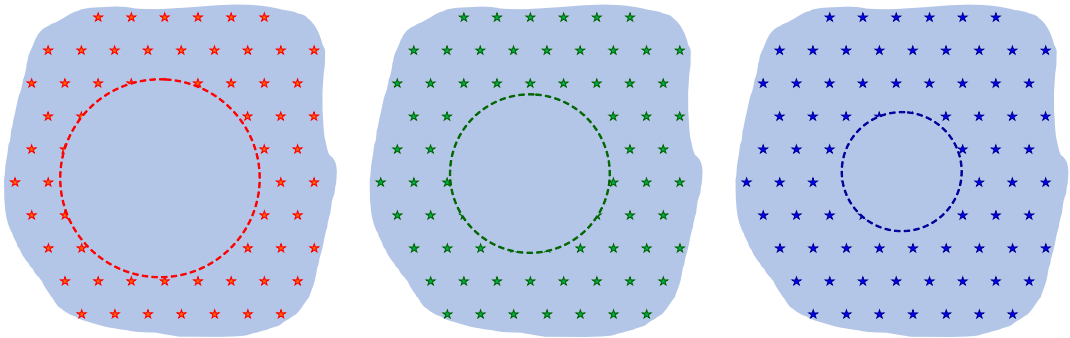}
	\caption{If a black hole is surrounded by a plasma, the observed angular size of the shadow depends on the photon frequency. Note that the picture is schematic, the angular radius is not the result of the calculations. Red color has a longer wavelength than blue, so this picture corresponds to a change in the size of the shadow when a black hole is located in homogeneous plasma.}
\label{figure-rainbow-shadow}
\end{figure}

In the presence of a plasma the size of the shadow depends on the
wavelength at which the observation is made (Fig. \ref{figure-rainbow-shadow}), in contrast to the vacuum
case where it is the same for all wavelengths. Very high photon frequencies corresponds to vacuum case. The difference 
from the vacuum size of the shadow becomes bigger with decrease of the photon frequency $\omega_0$. Dependence of the shadow size on the photon frequency 
in a homogeneous plasma goes as 
\begin{equation}\label{eq:shadow-distant-non-hom}
\mbox{}{if} \quad \omega_0^{(1)} < \omega_0^{(2)} < \omega_0^{(3)}, \quad \mbox{then} \quad \alpha _{\mathrm{sh}} (\omega_0^{(1)}) > \alpha _{\mathrm{sh}} (\omega_0^{(2)}) > \alpha _{\mathrm{sh}} (\omega_0^{(3)}) > \alpha _{\mathrm{sh}}^{vacuum} \, ;
\end{equation}

\noindent and in a non-homogeneous plasma for a distant observer we have
\begin{equation}\label{eq:shadow-distant-non-hom}
\mbox{if} \quad \omega_0^{(1)} < \omega_0^{(2)} < \omega_0^{(3)}, \quad \mbox{then} \quad \alpha _{\mathrm{sh}} (\omega_0^{(1)}) < \alpha _{\mathrm{sh}} (\omega_0^{(2)}) < \alpha _{\mathrm{sh}} (\omega_0^{(3)}) < \alpha _{\mathrm{sh}}^{vacuum} \, .
\end{equation}

For estimation of the effect, the case of a spherically symmetric accretion of plasma onto a
Schwarzschild black hole was considered in detail. We have found that for an
observer far away from the Schwarzschild black hole the plasma
makes the shadow smaller. As examples, we have considered Sgr A* and M87.
Using this specific accretion model and observed luminosities in these systems, we have estimated the plasma density, and found that
the effect of the presence of a plasma on the size of the shadow
can be significant only for wavelengths of at least a few centimeters.
At such wavelengths the observation of the shadow is made difficult
because of scattering. For further details see Ref.~\cite{Perlick2015}.\\


\section{Influence of plasma on the shadow of a Kerr black hole}

As we learned from the last section, in the case of a spherically symmetric black hole, the shadow boundary is determined by the rays coming from the observer and falling on one of the circular orbits filling the photon sphere. The photon sphere is filled with circular orbits of the same radius.
In the case of the Kerr black hole, instead of the photon sphere, there is a so called photon region. The photon region is the region in spacetime filled with
spherical light rays, i.e., with solutions to the ray equation
that stay on a sphere $r =$ constant. Photon region includes spherical orbits of different radii. Similar to spherically symmetric case, unstable spherical
light rays can serve as limit curves for light rays that
approach them in a spiral motion.

\begin{figure}
	\centering
	\includegraphics[width=12cm]{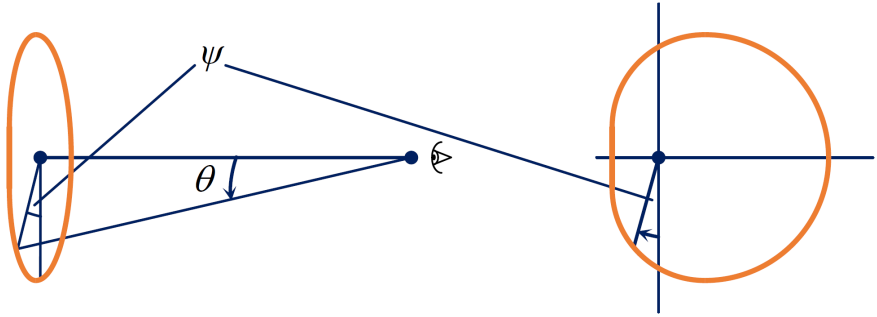}
	\caption{Celestial angles in the observer's sky, for more details see Fig. 8 in \cite{GrenzebachPerlickLaemmerzahl2014}.}
	\label{figure-celestial-angles}
\end{figure}

The shadow in case of Kerr black hole is constructed as follows \cite{GrenzebachPerlickLaemmerzahl2014}. The observer sends the rays into the past. Some of them are absorbed by a black hole, others go to infinity. The first type of rays corresponds to the observer's darkness, the second type of rays corresponds to brightness. The boundary of the shadow is determined by the rays that asymptotically spiral towards one of the unstable spherical light orbits. For every azimuthal angle $\psi$ in the observer's sky there exist one spherical orbit with some radius $r_p$, which leads us to a value of colatitude angle $\theta$, see Fig.\ref{figure-celestial-angles}. The set of these orbits determines the shadow of a Kerr black hole. Change of the azimuthal angle corresponds to a change of the parameter $r_p$ from some minimal to some maximum value. Using this way of construction, we can obtain the shadow written in terms of two celestial angles.

In this Section we review results of recent paper of Perlick and Tsupko \cite{PerlickTsupko2017}. Light propagation in a non-magnetized pressureless plasma
around a Kerr black hole is considered. We have found the necessary and sufficient condition
the plasma electron density has to satisfy, to guarantee that the
Hamilton-Jacobi equation for the light rays is separable, i.e., that
a generalized Carter constant exists. The shadow of a Kerr
black hole under the influence of a plasma, that satisfies the separability
condition is calculated. More precisely, an analytical formula for the
boundary curve of the shadow on the sky of an observer that is located
anywhere in the domain of outer communication is derived.

We consider the Kerr metric, in Boyer-Lindquist
coordinates $x=(t, r, \vartheta , \varphi )$,
\begin{equation}
ds^2  =
-c^2  \left(1-\frac{2mr}{\rho^2}\right) dt^2 +
\frac{\rho^2}{\Delta} dr^2 + \rho^2 d\vartheta^2 \, + 
\end{equation}
\[
+ \, \mathrm{sin}^2\vartheta\left(r^2+a^2+
\frac{2mra^2\mathrm{sin}^2\vartheta}{\rho^2}\right)
d\varphi^2 - \frac{4mra \mathrm{sin} ^2\vartheta}{\rho^2} \, c \, dt \, d\varphi
\]
where
\begin{equation}\label{eq:Deltarho}
\Delta = r^2 + a^2 - 2mr \, , \quad
\rho ^2 = r^2 + a^2 \mathrm{cos} ^2 \vartheta \, .
\end{equation}
Here $m$ is the mass parameter and $a$ is the spin parameter,
\begin{equation}\label{eq:mM}
m = \frac{GM}{c^2} \, , \quad a = \frac{J}{Mc} \, ,
\end{equation}
where $M$ is the mass and $J$ is the spin.
We assume that the plasma electron frequency $\omega_e$ is a function only of $r$ and $\vartheta$.

We have used the Hamiltonian formalism for light rays in a plasma on
Kerr spacetime and have determined the necessary and sufficient condition for separability of
the Hamilton-Jacobi equation.
We have demonstrated that the Hamilton-Jacobi equation is separable, i.e., that a
generalized Carter constant exists, only for special distributions of the
plasma electron density:
\begin{equation}\label{eq:sepcon}
\omega^2_e (r , \vartheta )  = \frac{f_r(r)+f_{\vartheta} ( \vartheta )}{
	r^2 + a^2 \mathrm{cos} ^2 \vartheta}
\end{equation}
with some functions $f_r(r)$ and $f_{\vartheta} ( \vartheta )$.

We have derived analytical formulas for the boundary curve of the shadow on
the observer’s sky in terms of two angular celestial coordinates, colatitude $\theta$ and azimuthal $\psi$ angles:
\begin{equation}\label{eq:shtheta}
\mathrm{sin} \, \theta =
\left.
\sqrt{\frac{\big(K-f_{\vartheta} ( \vartheta ) \big) \Delta}{
		\Big( (r^2+a^2)\, \omega _0 + a \, p_{\varphi} \Big)^2
		- \Big( f_r(r)+f_{\vartheta} ( \vartheta ) \Big) \Delta }}
\right|_{(r_{\mathrm{O}},\vartheta_{\mathrm{O}})}  \, ,
\end{equation}
\begin{equation}\label{eq:shpsi}
\mathrm{sin} \, \psi = \left.
\frac{ - p_{\varphi} - a \, \mathrm{sin} ^2 \vartheta \, \omega _0 }{
	\mathrm{sin} \, \vartheta \, \sqrt{K-f_{\vartheta} ( \vartheta )}}
\right|_{\vartheta_{\mathrm{O}}}  \, ,
\end{equation}
with the constants of
motion $a p_{\varphi}$ and $K$,
\begin{equation}\label{eq:pphispher}
a p_{\varphi}  = \, \frac{\omega _0}{(r-m)}
\left( m(a^2-r^2) \pm
r \Delta \sqrt{1-f_r'(r) \frac{(r-m)}{2 r^2 \omega _0 ^2}} \right)
\, ,
\end{equation}
\begin{equation}\label{eq:Kspher}
K = \frac{r^2 \Delta \omega _0 ^2}{(r-m)^2}
\left(1 \pm \sqrt{1-f_r'(r) \frac{(r-m)}{2 r^2 \omega _0 ^2}} \right) ^2
-f_r(r) \, .
\end{equation}
With these expressions for $K(r_p)$ and $p_{\varphi}(r_p)$ inserted
into (\ref{eq:shtheta}) and (\ref{eq:shpsi}) we get the boundary
of the shadow as a curve on the observer's sky parametrized by parameter $r_p$.
The boundary curve consists of a lower part, where $\psi$ runs from
$- \pi/2$ to $\pi /2$, and an upper part where $\psi$ runs from $\pi /2$
to $3 \pi /2$. The parameter $r_p$ runs from a minimum value
$r_{p, \mathrm{min}}$ to a maximum value $r_{p, \mathrm{max}}$
and then back to $r_{p, \mathrm{min}}$. The values $r_{p, \mathrm{min}}$
and $r_{p, \mathrm{max}}$ are determined by the property that then
$\mathrm{sin} ^2 \psi$ must be equal to 1, i.e., by (\ref{eq:shpsi}),
$r_{p} = r_{p, \mathrm{min}/\mathrm{max}}$ if
\begin{equation}\label{eq:rpminmax}
- p_{\varphi} (r_p ) - \, a \, \mathrm{sin} ^2 \vartheta _{\mathrm{O}}
\, \omega _0
\, = \, \pm \,
\mathrm{sin} \, \vartheta _{\mathrm{O}}
\, \sqrt{K(r_p)-f_{\vartheta } ( \vartheta _{\mathrm{O}})}
\, .
\end{equation}
Our formulas are valid for any photon frequency at infinity $\omega _0$, for any value
of the spin parameter $a$, for any position of the observer outside of the horizon
of the black hole (i.e., with arbitrary distance from the black hole and arbitrary
inclination) and for any plasma distribution which satisfies the separability
condition.

\begin{figure}[h]
	\centering
	\includegraphics[width=9cm]{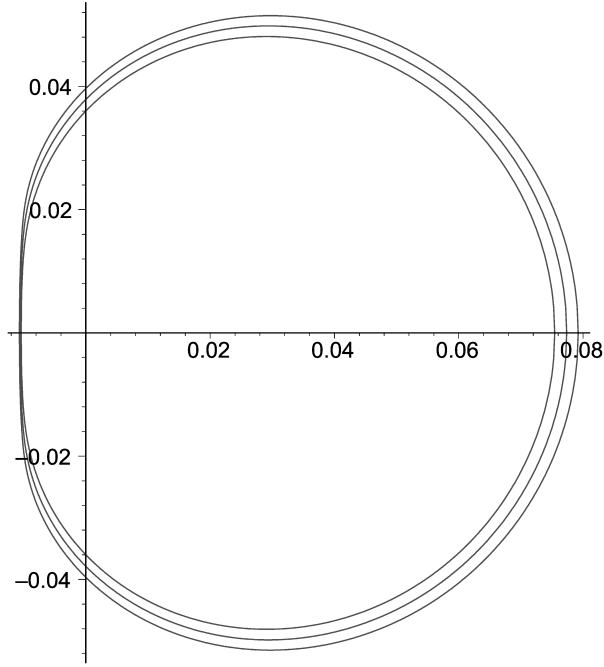}
	\caption{Shadow for a distant equatorial observer for $f_r(r)= \omega _c ^2 \sqrt{m^3r}$
and $f_\vartheta(\vartheta)=0$ with a constant $\omega _c$. The parameters are
$a=0.999m$, $r_\mathrm{O}=100m$, $\vartheta_\mathrm{O}= \pi /2$.
The three curves correspond to $\omega _0 =\infty$ (outermost curve),
$\omega _0 = \omega _c$ and $\omega _0 = \omega _c /\sqrt{2}$.}
	\label{figure-shadow-kerr}
\end{figure}

When plotting the shadow (for example, see Fig. \ref{figure-shadow-kerr}), we use
stereographic projection (see Fig. 8 in \cite{GrenzebachPerlickLaemmerzahl2014}) onto a plane that is tangent to the celestial
sphere at the pole $\theta =0$, and in this plane we use dimensionless
Cartesian coordinates,
\begin{equation}
\label{eq:stereo}
X(r_p) = -2 \tan \Big( \frac{\theta(r_p)}{2} \Big)
\sin \big( \psi(r_p) \big) \, , \quad Y(r_p) = -2 \tan \Big( \frac{\theta(r_p)}{2} \Big)
\cos \big( \psi(r_p) \big).
\end{equation}
The pole $\theta =0$, i.e., the origin of our Cartesian 
coordinate system, corresponds to a past-oriented ingoing
principal null ray through the observer position.
This method of plotting the shadow, which was used in
\cite{GrenzebachPerlickLaemmerzahl2014}, \cite{GrenzebachPerlickLaemmerzahl2015}, \cite{PerlickTsupko2017}, is to be distinguished
from the one introduced by Bardeen \cite{Bardeen1973} and used subsequently by many authors.
Bardeen considers an observer at infinity with use of impact parameters as the coordinates on the axes, while the presented approach allow any
observer position in the domain of outer communication, and coordinates are directly related to 
angular measures on the observer's sky. Also, origins of coordinates in these approaches have different positions (of course, the choice of the origin is a matter of convention). One just has to keep in mind that presented way of plotting
may be directly compared with Bardeen's only if the observer is at a big radius
coordinate and that the origin is horizontally shifted, for details see text after formula (57) in paper \cite{PerlickTsupko2017}.

For the reader's convenience, we have written a step-by-step procedure
for the construction of the shadow \cite{PerlickTsupko2017}.

1. Choose the mass parameter $m$ and the spin parameter $a$ with $a^2 \le m^2$.
The mass parameter $m$ gives a natural length unit, i.e., all other lengths may be given
in units of $m$.

2. Choose a plasma frequency $\omega _e ( r , \vartheta )$ around the black hole.
The plasma frequency has to satisfy the separability condition (\ref{eq:sepcon}), so
the plasma distribution is characterized by two functions $f_r(r)$ and
$f_{\vartheta} ( \vartheta )$. Only for such plasma frequencies
are the ray equations completely integrable and the shadow can be calculated
analytically.

3. Choose a position of an observer anywhere in the domain of outer communication
by prescribing its radial and angular coordinate $r_\mathrm{O}$ and $\vartheta_\mathrm{O}$.
For an illustration see Fig. 7 in \cite{GrenzebachPerlickLaemmerzahl2014}.

4. Choose the constant of motion $\omega_0$ for the rays that are to be considered.
In the formulas for the shadow, $\omega _0$ will enter only in terms of the quotient
$\omega^2 _e (r , \vartheta)^2/\omega _0$. Therefore, it is convenient to give
$\omega _e (r , \vartheta )$ and $\omega _0$ as multiples of the same frequency
unit $\omega _c$ which will then drop out from all relevant formulas.

5. Write the celestial coordinates $\sin \theta$ and $\sin \psi$ in terms of the
constants of motion of the corresponding ray by
(\ref{eq:shtheta}) and (\ref{eq:shpsi}) with $r=r_\mathrm{O}$ and
$\vartheta=\vartheta_\mathrm{O}$ substituted. For an illustration of the
angles $\theta$ and $\psi$ see Fig. \ref{figure-celestial-angles} of present paper and Fig. 8 in \cite{GrenzebachPerlickLaemmerzahl2014}.

6. Substitute into the expressions for $\sin \theta$ and $\sin \psi$
the expressions $K(r_p)$ and $p_\varphi(r_p)$ according to formulas (\ref{eq:pphispher})
and (\ref{eq:Kspher}) with $r=r_p$. Here $r_p$ runs over an interval of radius
coordinates for which unstable spherical light rays exist. This gives us $\sin \theta$ and
$\sin \psi$ as functions of  $r_p$, i.e., it gives us a curve on the observer's
sky. This is the boundary curve of the shadow. In the next two steps we determine
the range of the curve parameter $r_p$.

7. Solve the equation
$\sin \psi(r_p) = 1$ for $r_p$. This gives us the minimal value $r_{p, \mathrm{min}}$. If the plasma density is small and if
the observer is in the equatorial plane, for small
$a$ it will be $r_{p, \mathrm{min}} \lesssim 3m$, while for a nearly extreme Kerr
black hole ($a \lesssim m$) it will be $r_{p, \mathrm{min}} \gtrsim m$.

8. Solve the equation $\sin \psi(r_p) = -1$ for $r_p$. This gives us the maximal
value $r_{p, \mathrm{max}}$. If the plasma density is small,
for small $a$ it will be
$r_{p, \mathrm{max}} \gtrsim 3m$, while for a nearly extreme Kerr black hole it
will be $r_{p, \mathrm{max}} \lesssim 4m$.

9. Calculate $\sin \theta(r_p)$ and $\sin \psi(r_p)$ where $r_p$ ranges over the
interval $\,] \, r_{p, \mathrm{min}} , r_{p, \mathrm{max}} \, [ \,$.
Note that $\theta(r_{p, \mathrm{min}}) +
\theta(r_{p, \mathrm{max}})$ gives the horizontal diameter of the shadow.

10. Calculate the dimensionless Cartesian coordinates $X$ and $Y$ of the boundary curve
of the shadow by formulas (\ref{eq:stereo}). Choosing $-\pi /2 \le \psi (r_p) \le \pi /2$
and letting $r_p$ run from $r_{p, \mathrm{min}}$ to $r_{p, \mathrm{max}}$ gives the
lower half of the boundary curve of the shadow. The upper half of the curve is the
mirror image of the lower half with respect to a horizontal axis.\\

\section{Conclusions}

(i) Assuming that the influence of both gravity and plasma on the trajectory is small, the total deflection angle 
can be calculated as the sum of individual effects: gravitational deflection and refraction, both angles are small.
 Effects of deflection by gravity in vacuum and the refractive deflection in non-homogeneous medium are well known
  and presented in literature with necessary details.

(ii) In the case of a strong gravitational field and for the study of subtle effects, a self-consistent theory of 
gravitation in matter is necessary. In presence of both gravity and plasma the deflection angle is physically defined 
by mutual combination of different phenomena: gravity, dispersion, refraction.

(iii) New effect is that in the case of a homogeneous plasma, in absence of refractive deflection, the gravitational 
deflection itself differs from the vacuum deflection and depends on the photon frequency. Presence of plasma makes 
the gravitational deflection bigger.

(iv) Plasma is a dispersive medium, its refractive index depends on the photon frequency. Therefore the deflection 
angle in presence of both gravity and plasma always depends on the photon frequency. In particular, presence of 
homogeneous plasma leads to chromatic gravitational deflection; the presence of non-homogeneous plasma leads, in 
addition  to the chromatic gravitational deflection, to a chromatic refractive deflection. Refractive deflection on 
the density profile falling radially outwards is opposite to the chromatic gravitational deflection by central mass.

(v) Presence of plasma makes gravitational lensing chromatic and leads to difference in angular position of the same 
image at different wavelengths. With present level of technique, the chromatic plasma effects might be observed only
 in strong lens systems. All chromatic effects are significant only for very long radiowaves.

(vi) The deflection angles for light rays near Schwarzschild black hole in homogeneous plasma, in strong deflection 
limit, are derived. Using of strong deflection limit allows us to investigate analytically properties of relativistic 
images in presence of plasma.

(vii) Analytical formula for the angular size of the shadow in a plasma with
a spherically symmetric density distribution on a spherically
symmetric and static spacetime is derived. In the presence of a plasma the size of the black hole shadow depends 
on the wavelength at which the observation is made, in contrast to the vacuum case where it is the same for all 
wavelengths. In homogeneous plasma, the shadow becomes bigger for distant observer, in comparison with vacuum case.
 In non-homogeneous plasma, the shadow becomes smaller for distant observer.
As examples, we have considered Sgr A* and M87, and found that
the effect of the presence of a plasma on the size of the shadow
can be significant only for wavelengths of at least a few centimeters (radio regime).

(viii) The propagation
of light rays in a non-magnetized pressureless plasma
on Kerr spacetime is investigated. We have
demonstrated that the Hamilton-Jacobi equation is separable,
i.e., that a generalized Carter constant exists, only
for special distributions of the plasma electron density. We have derived analytical formulas for the boundary
curve of the shadow on the observers sky in terms of two
angular celestial coordinates.

\section*{Acknowledgments}
This work was supported by Russian Science Foundation, Grant No. 15-12-30016. We would like to thank our collaborator and friend Volker Perlick for many valuable discussions, especially for introducing us to the subject of black hole shadow. We are also thankful to the four anonymous Referees for their extremely careful reading of the present manuscript and a number of valuable comments.






\end{document}